%% file: yodel.tex
\documentclass{article}

\pdfoutput=1 

\usepackage{arxiv}

\usepackage[utf8]{inputenc} 
\usepackage[T1]{fontenc}    
\usepackage{booktabs}       
\usepackage{amsfonts}       
\usepackage{nicefrac}       
\usepackage{microtype}      
\usepackage{lipsum}         
\usepackage{graphicx}
\usepackage[square,numbers]{natbib}
\usepackage{dirtytalk}
\usepackage{enumitem}
\usepackage[dvipsnames]{xcolor}
\usepackage{tabularray}
\usepackage{animate}

\title{Yodel: A Layer 3.5 Name-Based Multicast Network Architecture For The Future Internet}

\author{ \hspace{1mm}\color{RedViolet}Morteza Moghaddassian \thanks{\footnotesize This work has been submitted to the IEEE for possible publication. Copyright may be transferred without notice, after which this version may no longer be accessible.}
and \hspace{1mm}Alberto Leon-Garcia \vspace{3mm}\\
	\small The Edward S. Rogers Sr. Department of Electrical and Computer Engineering \\
        \small University of Toronto, ON M5S 3G4, Canada \\
	\texttt{\small \{m.moghaddassian,alberto.leongarcia\}@utoronto.ca} \\
}

\begin{document}
\maketitle

\input{content/abstract}

\color{RedViolet} 

\input{content/introduction}
\input{content/literature}
\input{content/goals}
\input{content/architecture}
\input{content/rf}
\input{content/dt}
\input{content/summary}
\input{content/media}
\input{content/naming}

\newpage

\appendix
\input{content/service}

\bibliographystyle{unsrtnat}
\bibliography{yodel}

\end{document}

%% file: content/abstract.tex
\color{RedViolet} \begin{abstract} \color{Black}

Multicasting refers to the ability of transmitting data to multiple recipients without data sources needing to provide more than one copy of the data to the network. The network takes responsibility to route and deliver a copy of each data to every intended recipient. Multicasting has the potential to improve the network efficiency and performance (e.g., throughput and latency) through transferring fewer bits in communicating the same data to multiple recipients compared with unicast transmissions, reduce the amount of networking resources needed for communication, lower the network energy footprint, and alleviate the occurrence of congestion in the network. Over the past few decades, providing multicast services has been a real challenge for ISPs, especially to support home users and multi-domain network applications, leading to the emergence of complex application-level solutions. These solutions like Content Delivery and Peer-to-Peer networks take advantage of complex caching, routing, transport, and topology management systems which put heavy strains on the underlying Internet infrastructures to offer multicasting services. In reality, the main motivation behind the design of these systems is rather sharing content than offering efficient multicast services. In this paper, we propound Yodel, a name-based multicast network architecture that can provide multi-domain multicast services for current and future Internet applications. Compared to the wider array of other name-based network architectures with clean-slate infrastructure requirements, Yodel is designed to provide multicast services over the current Internet infrastructure. Hence, Yodel puts forward several design goals that distinguish it from other name-based network architectures with inherent multicast capabilities. This paper is prepared to discuss the Yodel architecture, its design goals, and architectural functions. 

\end{abstract}

%% file: content/introduction.tex
\color{RedViolet} \section{Introduction} \color{Black}

The current Internet architecture was initially designed to facilitate a basic one-to-one host-centric communication model and has since scaled successfully to meet the requirements of the Internet unicast applications such as email, banking, and file transfer. With the advent of Web and Mobile platforms, however, new Internet applications have emerged that are inherently based on multicast communication models of one-to-many (e.g., podcasts and audio/video streaming) and many-to-many (e.g., group messaging, remote conferencing, and multiplayer games). These applications could not initially rely on network-level support to meet their multicast needs, thereby were obliged to use unicast communications to achieve multicast functionality. Unfortunately, using unicast connections to build multicasting began to put a strain on the Internet networking infrastructure as it began to grow globally which consequently leading to massive growth in the number of Internet users and applications.

Of course, the need for efficient support of multicast applications did not go unnoticed and led to the emergence of several different proposals for implementing multicast at large scales. Perhaps, the first to emerge has been the IP Multicast architecture \cite{rfc5110-savola-IP_Multicast, IP_Multicast-deering-1994}. IP Multicast has exclusively designed to enable end-to-end multicast services for the current IP-based Internet architecture. In particular, an IP Multicast network can enable one or more data sources in an IP network to send their data to a group of recipients where the senders are only required to transmit one copy of their data to the network \cite{rfc5110-savola-IP_Multicast, IP_Multicast-deering-1994}. The IP Multicast network can subsequently deliver a copy of the senders data to every intended recipient across a multi-domain network infrastructure. In this scenario, the sender(s) and receivers must have already joined a pre-established multicast group for which IP Multicast uses class D IP addresses \cite{rfc5110-savola-IP_Multicast}. Given the IP Multicast Internet-friendly approach of using existing Internet networking infrastructures \cite{rfc5110-savola-IP_Multicast, IP_Multicast-deering-1994}, the architecture has been widely considered as a viable solution for adding multicast to the current Internet architecture. However, several decades of researching the key IP Multicast challenges like addressing, scalability, routing, group management, and security have not successfully incentivesed widespread deployment of the technology, especially for supporting home users. To fill the gap, alternative approaches have gradually emerged that either stem from the classic IP Multicast like IPv6 \cite{IPv6-stalling-1996, rfc8200-deering-IPv6} or methods that rely on overlays and persistent tunnels to reduce topology changes and the need for constant rerouting in multicast networks like the MBone architecture \cite{Mbone-eriksson-1994}. However, these methods are also not widely used in today's Internet architecture mainly for the same reasons as IP Multicast does not and the rise of alternative application-level solutions like the Content Delivery \cite{CDN_intro-zolfaghari-2020} and Peer-to-Peer Networks \cite{P2P_intro-pourebrahimi-2005}.

Today, application-level multicast systems address a large sum of the Internet's multicast requirements. These systems typically maintain directories of applications data names and methods of mapping them to network addresses/locations (e.g., DNS) to provide alternative methods of distributing applications data to the Internet users over the current Internet unicast-based Infrastructure \cite{multicast_survey_application-yeo-2004,multicast_survey_application-hosseini-2007}. In contrast to IP Multicast, these alternative methods have demonstrated cost-effectiveness (i.e., no infrastructure changes is needed) and greater adaptability to the changing demands of the Internet users in multicast network environments \cite{multicast_survey_application-yeo-2004,multicast_survey_application-hosseini-2007}. Nonetheless, utilizing complex protocol stacks and advanced techniques in caching, routing, and topology management used by these systems for reliability and efficiency can place significant demands on the existing Internet infrastructures. Another approach is to directly use names at the network level. Name-based networking is proposed by Information-Centric Networking (ICN) \cite{ICN_survey-boutaba-2012, ICN_survey-ahlgren-2012} as an alternative method for building the Internet architecture \cite{fia-nsf-2024}. Basically, instead of using network addresses, a name-based network uses data names to identify and move data objects in the network. In particular, every distinct data object in the network can be published to and requested by its unique name where the sender(s) and receivers of the data can inherently form a multicast group that is tied to the distinct data object name \cite{ICN_survey-boutaba-2012, ICN_survey-ahlgren-2012}. In essence, name-based networking can offer multicast functionalities akin to application-level solutions but at the network level. 

We emphasize that the network-level support is preferable for multicast applications, as it eliminates dependencies on external systems and removes the incogruity between using names at the application level and the underlay host-centric network infrastructure. Names also decouple data from a particular sender/location in the network, thus have the potential to simplify multicast transmissions and group memberships management specially in mobile networks \cite{ICN_mobility-fayyaz-2023} and multi-domain settings as data objects can be selectively sent from any (e.g., nearest, less loaded, etc.) data source/location in the network \cite{ICN_survey-boutaba-2012, ICN_survey-ahlgren-2012}. One might contend that a comparable functionality (i.e., choosing any available data source) can be also attained using names at the application level and even with IP Multicast addresses. However, we wish to underscore the complexity of application-level systems and the expected intricacy associated with adapting the IP Multicast signaling and routing algorithms to accommodate this requirement, in contrast to the seamless and inherent support achievable through name-based networking \cite{ICN_survey-boutaba-2012, ICN_survey-ahlgren-2012}.

Given the attainable network-level support and effective multicast capabilities that are inherent in name-based networking, particularly when compared to the broader landscape of other multicast solutions, we believe that the name-based networking can stand as a promising candidate for meeting the multicast requirements of current and future Internet applications. Of course, there are key existing challenges in designing and successfully implementing name-based networks for providing multicasting services to the Internet applications such as infrastructural incompatibility, routing scalability, and interdomain routing as well as the heterogeneity of multicast applications needs that require thoughtful considerations. To meet these challenges, we propound the Yodel architecture \cite{yodel-morteza-2023}. Compared to the wider array of other name-based network architectures, Yodel incorporates specific design goals that allow the architecture to address the aforementioned challenges in such a way that is particularly well-suited for the use in multicast application environments. Generally, existing name-based network architectures are typically proposed as full-fledged network architectures aiming to replace the current Internet architecture and networking infrastructures \cite{fia-nsf-2024, ICN_survey-boutaba-2012, ICN_survey-ahlgren-2012}. In contrast, Yodel is not a clean-slate proposal but rather is an Internet-friendly approach for enabling name-based multicast services using a layer 3.5 approach that makes it compatible with current Internet networking infrastructures while also enabling it to support name-based multi-domain routing and forwarding. In addition to infrastructural compatibility, Yodel also benefits Internet multicast applications in several other ways, including inherent support for multiple multicast service models, resiliency against temporary network disconnections, and multi-tenancy that we will explain in details in section 3 and throughout this paper. 

The rest of this paper continues by first providing a summary background and review of existing approaches for building network-level multicast services in section 2. In section 3, we identifying the Yodel design goals and contributions. In section 4, we describe the Yodel architecture, its key elements and functions, and multicast service models. section 5 provides a detailed discussion of routing, multicast group management, and forwarding in Yodel. In section 6, we discuss connection resiliency and describe how Yodel realizes that as a way to improve the performance of multicast applications in various networking environments. Finally, we conclude this paper by providing a summary and a discussion of the future works in section 7. This paper also include some sections that explain additional information about Yodel and this paper, including a detailed appendix section that describe Yodel Multicast services in more details and provides use cases that can help readers to better understand the use of Yodel multicast services in different application settings.

%% file: content/literature.tex
\color{RedViolet} \section{Background And Related Work}\color{Black}

As we discussed in section 1, multicasting can be offered in both application and network level capacities. Some networking technologies like Multi Protocol Label Switching (MPLS) and Line Speed Publish/Subscribe Inter-networking (LIPSIN) \cite{lipsin-jokela-2009} can also provide multicasting services at layers 2/2.5. Since the body of work is huge and historically deep, and the focus of this paper and the Yodel architecture is also on providing name-based network-level multicast services, in this section we only focus on reviewing related works from the perspective of network-level multicast solutions with an emphasis on IP Multicast and ICN methodologies. The aim is not to build a comprehensive architectural comparison of current and the future Internet architectures but rather to focus on the multicast design approaches from the perspective of dissimilar architectural viewpoints (i.e., host-centric versus name-based) so that we can build a foundation for better describing Yodel and its design goals to our valued readers in the rest of this paper. 

\color{DarkOrchid} \subsection{IP Multicast Design Space}\color{Black}

The Internet Protocol Multicast architecture, also known as IP Multicast, was first introduced and standardized in 1986 \cite{rfc1112-deering-1989} with the aim of enabling data sources in an IP network to simultaneously transmit data packets to multiple recipients \cite{rfc1112-deering-1989, IP_Multicast-deering-1994} without needing to send more than one copy of each data packet to the network. To do so, all senders and intended recipients of a multicast communication must first join a pre-established multicast group that is identified by a class D IP address for IPv4 networks or a ff00::/8 address for IPv6 networks. Data sources in the group can then send their data to the group's recipients in IP packets that contain the source (unicast) IP address and the multicast group's address as the intended destination. The IP Multicast architecture can then deliver data packets to the group's recipients using two distinct service model of: 1) Any Source Multicast (ASM), and 2) Source Specific Multicast (SSM) models. The Any Source Multicast service model \cite{rfc5110-savola-IP_Multicast} which has also been the default multicast service model during the early years of IP multicast deployment \cite{asm-maufer-1998} allows more than one data sources in the group to send their data packets to the groups recipients while the Source Specific Multicast service model \cite{ssm-bhat-2003, rfc5110-savola-IP_Multicast} allows recipients in the multicast group to receive data from a single and specific data source in the group. To realize both service model, the IP Multicast architecture relies on a series of protocols and algorithms for handling routing and forwarding, source discovery, and group management that we will review in the next sections.

\color{DarkOrchid}\subsubsection{Group Management In IP Multicast}\color{Black}

Group management in IP Multicast is handled by IGMP \cite{IP_Multicast-ratnasamy-2006} and MLD \cite{IP_Multicast-ratnasamy-2006} family of backwards-compatible protocols known as IGMPv1 \cite{rfc1112-deering-1989}, IGMPv2 \cite{igmp_2-fenner-1997}, and IGMPv3 \cite{igmp_3-cain-2002} in IPv4 networks and on MLDv1 \cite{rfc2710-deering-1999} and MLDv2 \cite{rfc3810-vida-2004} protocols in IPv6 networks. Generally, IGMP/MLD protocols operate between IP-based hosts (i.e., client/user devices) on a multicast network and the network's IP multicast routers for handling multicast group management activities  \cite{IP_Multicast-ratnasamy-2006} like adding, deleting, and updating hosts registration in different multicast groups (i.e., addresses). To do so, IGMP/MLD protocols first enable hosts and IP multicast routers in the network to communicate and exchange group registration and memberships information which will be later used for source discovery and routing purposes. We note that not all routers in a network are multicast-enabled routers. Second, IGMP/MLD protocols enable the network switches in between the hosts and the IP multicast routers to perform IGMP snooping \cite{igmp_3-cain-2002}, a process that enables switches in a local network to listen to IGMP messages and maintain state information about the hosts registrations in different multicast groups (i.e., addresses). Such information is later used by the network switches to forward groups data packets towards intended recipients in the local network. To send data packets across the networks, IP Multicast routers between the source and destination hosts (i.e., multicast group recipients) must use group memberships information for performing source discovery and routing to create multicast forwarding trees.

\newpage

\color{DarkOrchid}\subsubsection{Routing In IP Multicast}\color{Black}

Routing in IP Multicast networks is performed using Protocol Independent Multicast (PIM) \cite{PIM-deering-1996, PIM_motivation-deering-1995}. PIM refers to a family of algorithms that enable one-to-many and many-to-many delivery of data packets to the recipients of a multicast group. The PIM protocol family includes three key variants for supporting Any Source Multicast service model. 

\begin{itemize}[leftmargin=0.3cm]
    \item \textbf{PIM-SM}: The PIM Sparse Mode (PIM-SM) enables the creation of unidirectional forwarding trees for separate multicast groups in the network \cite{pim_sm_motivation-deering-1996, pim_sm_spec-estrin-1998}. The protocol uses Rendezvous Points (RP) for performing end-to-end routing of multicast data packets. Within a multicast network, any of the available multicast routers can act as the network initial rendezvous point. Other multicast routers can become a rendezvous point by joining the initial RP \cite{pim_sm_motivation-deering-1996, pim_sm_spec-estrin-1998}. Once a host sends a group registration request to a multicast router, the router records the incoming interface for the request and periodically sends updates regarding the registration status to other multicast routers in the network. Other multicast routers also forward this information until all rendezvous points for the multicast group will receive the registration updates \cite{pim_sm_motivation-deering-1996, pim_sm_spec-estrin-1998}. When a sender transmits a data packet to a multicast group, multicast routers use the Reverse Path Forwarding (RPF) (i.e., calculated by recording incoming interfaces for the group registration and update requests) to deliver the packet to the groups recipients. Multicast routers and rendezvous points also use topology information for source discovery and building multicast forwarding trees. When a host leaves a multicast group, the local rendezvous point sends prune updates to all other multicast routers in the network to avoid unnecessary packet delivery to the unsubscribed host \cite{pim_sm_motivation-deering-1996, pim_sm_spec-estrin-1998}. 
    
    \item \textbf{PIM-DM}: The PIM Dense Mode (PIM-DM) is another PIM family variant that creates unidirectional forwarding trees for separate multicast groups in the network \cite{pim_dm_spec-adams-2005}. PIM-DM major difference with PIM-SM \cite{pim_sm_motivation-deering-1996, pim_sm_spec-estrin-1998} is in its view to the network. While PIM-SM basic assumption is that there are subscribers for a multicast group in some places in the network (i.e., sparsely distributed), the PIM-DM basic assumption is that there are subscribers for a multicast group in most places (if not all) in the network (i.e., densely distributed) \cite{pim_dm_spec-adams-2005}. Such assumption allows the local multicast routers in a PIM-DM network to flood the hosts registrations information of different multicast groups to all neighbor multicast routers where each neighbor multicast router also repeats the process until all multicast routers receive the registrations information and become a rendezvous point for the multicast groups. If a multicast router shows no interest to participate as a rendezvous point for a multicast group, the router sends prune messages in the reverse forwarding path to update upstream multicast routers of its decision \cite{pim_dm_spec-adams-2005}. The upstream multicast routers use the prune messages to adjust the forwarding paths. The same as in PIM-SM \cite{pim_sm_motivation-deering-1996, pim_sm_spec-estrin-1998}, when a host leaves a multicast group in PIM-DM approach, the local rendezvous point sends prune updates to all other multicast routers in the network to avoid unnecessary packet delivery to the unsubscribed host \cite{pim_dm_spec-adams-2005}. 

    \item \textbf{Bidir-PIM}: The third variant in the PIM protocol family is known as Bidirectional PIM (Bidir-PIM) \cite{rfc5015_pim_bid-handley-2007}. Unlike the PIM-SM and PIM-DM protocols that construct unidirectional forwarding paths, the Bidir-PIM creates bidirectional multicast trees \cite{rfc5015_pim_bid-handley-2007}. The protocol uses the same assumption that PIM-SM uses to create multicast forwarding paths but it cannot always create a shortest path between the senders and the receivers in a multicast group.
\end{itemize}

IP Multicast can also support Source Specific Multicast service model. In this multicast model, recipients can register for a multicast group and can specify a source address with their registration request to receive the group's data from. Doing so also requires the recipients to know the given source/sender's identity (e.g., unicast IP address) to register their request. IP Multicast uses a variant of the PIM-SM protocol to handle source specific multicast requests, called the PIM Source Specific Multicast (PIM-SSM) \cite{ssm-bhat-2003}. Once a (sender) host registers in a multicast group, the local multicast router/rendezvous point also records the sender’s identity besides the group multicast address \cite{ssm-bhat-2003}. The rendezvous point, the same as in PIM-SM \cite{pim_sm_motivation-deering-1996, pim_sm_spec-estrin-1998}, informs all other rendezvous points in the multicast group of the host registration information. Once the sender transmits a data packet to the multicast group address, the packet will be only delivered to the recipients in the group that have registered for the given sender in the multicast group \cite{pim_sm_motivation-deering-1996, pim_sm_spec-estrin-1998}. We note that since the group rendezvous points know about the group data sources and addresses, there is no need for source discovery. We also note that IP Multicast can support multi-domain multicast routing using protocols like MSDP \cite{rfc3618_msdp-fenner-2003} and MBGP \cite{rfc4760_MBGP-bates-2007} regardless of the service model.

\color{DarkOrchid}\subsubsection{The MBone Proposal}\color{Black}

As stated in section I, despite the interesting design and infrastructural compatibility between the Internet and IP multicast architectures, ISP's today, specially smaller ones with less technical and executive strength, do not fully support IP Multicast capabilities. Partly, due to complex configurations that require experts and skilled technicians to manage the infrastructure, but also because of secondary challenges regarding billing and tracking bandwidth consumption which are not the target of these paper. To resolve the lack of infrastructure and ISP's presence in building a globally connected Internet multicast infrastructure, the Multicast Backbone (i.e., MBone) has emerged building a virtual multicast network atop the existing Internet infrastructure for carrying IP Multicast traffic \cite{Mbone-eriksson-1994}. The aim is to use the Internet architecture to achieve a global multicast network. To do so, MBone uses IP Multicast routers as rendezvous points \cite{Mbone-eriksson-1994} in places where IP Multicast is deployed. However, in places where an IP Multicast router is not accessible (lack of infrastructure support), MBone creates tunnels and carry multicast packets in unicast payloads \cite{Mbone-eriksson-1994}. The MBone network also uses Class D (224.4.2.0) IP address and IGMP for multicast group management but unlike IP Multicast, MBone uses Distance Vector Multicast Routing Protocol (DVMRP) for end-to-end routing of multicast data packets \cite{rfc1075_DVMRP-waitzman-1998}. The DVMRP is designed based on the Routing Information Protocol (RIP) \cite{rfc1058_RIP-hedrick-1988} and uses the number of multicast routers between a sender and a recipient in a multicast group as a parameter for creating multicast forwarding paths \cite{rfc1075_DVMRP-waitzman-1998}.

\color{DarkOrchid} \subsection{ICN Multicast Design Space}\color{Black}

As we discussed earlier, Information-Centric Networking (ICN) is a novel networking paradigm that aims to the use names instead of network addresses to uniquely identify and move data objects in the network. Names can generalize the notion of addresses in the network and build an abstract identity for the senders and receivers in the network that are interested in the same data object to communicate without recipients to specify the source or location of a given data object in the network to obtain the data nor sender(s) to program data packets for delivery to specific recipients in the network. In other words, senders of a given data object send their data to a name where the ICN network can deliver it to all the recipients in the network that have registered interests/subscriptions for the same data object name. Thus, ICN architectures can intrinsically support multicasting through naming. By analogy, names can be compared with multicast group addresses in IP Multicast. The key differences, however, are in the methods of addressing, name-data binding, and the type of protocols and algorithms used for routing, source discovery, and group management. In this section, we discuss routing, source discovery, and group management in the context of ICN networks. We will learn more about addressing and name-data binding in ICN networks in section 3.

\color{DarkOrchid}\subsubsection{The Binary (Duo) Messaging Model}\color{Black}

The binary (duo) messaging model refers to the method used by some ICN architectures like CCN \cite{ccn_intro-JJ-2007}, NDN \cite{ndn_intro-zhang-2014}, and DONA \cite{dona_intro-koponen-2007} wherein recipients in the network (i.e., data consumers) can issue Interest/Find messages that travel the network to find a matching data object. These messages carry the data object name in the message header. Once a data source/provider (i.e., data producer) is found in the network with a matching data object, the data producer can send a Data/Register message that contains the given data object and its data object name for delivery to the interested recipients. Although, these architectures are different in nature and the way they use Interest and Find messages for routing and source discovery, a prominent approach used by CCN/NDN architecture is that Data messages can be forwarded in the reverse forwarding path that the Interest messages have followed for delivery to the intended recipients \cite{ccn_intro-JJ-2007, ndn_intro-zhang-2014}. This approach can natively support multicasting in the network as the reverse forwarding path is a multicast tree. There are two key design decision pathways (among all) to realize the idea of using reverse forwarding paths for delivery of data objects to the intended recipients. 

In the first approach, every content/data router in a forwarding path adds its own router ID/outgoing interface address to the Interest messages header before moving them forward towards the matching data producers in the network so that the matching data producers can read the entire forwarding path stored in the Interest messages header and embed the reverse of the forwarding path in the Data messages header at the source. Every data/content router in a forwarding path can then remove its own ID/next interface address from the forwarding path stored in the Data messages header and move the Data messages forward until they are delivered to their destination. In this approach, content/data routers can become stateless as they do not need to hold forwarding tables. We note that routers send one Interest message forward for all the Interest messages they receive from data consumers that are interested in the same data object. However, for that to happen, routers may need to hold some state information to handle Interests aggregation. We also note that some variations of this approach is used in designing proprietary technologies in the context of ICN.

In the second approach, a local content/data router receive the Interest messages from data consumers in its locality. The router records the interfaces where every Interest message has arrived and send the Interest messages to its neighbor routers. We note that in this scenario, a single Interest message is always sent forward for all the Interest messages a router receives from all the data consumers that are interested in the same data object. In this way, once a matching data source/provider is found for the requested data object, the data source/provider can send the data object to the interested recipients in Data messages that only need to carry the name of the data object. Each intermediary content/data router in the forwarding path then uses that name to map the Data messages with the list of all downstream interfaces where the corresponding Interest messages for the given data object have arrived and forward the Data messages to all of them in parallel. This process continues until all Data messages are delivered to the intended recipients (i.e., interested data consumers). To perform routing and source discovery in both approaches, a link-state routing protocol can be also used \cite{nlsr-hoque-2013}, as suggested by NDN proposal \cite{ndn_intro-zhang-2014}. In the case of DONA, a multi-tier rendezvous system (i.e., resolution handling system) is used instead \cite{dona_intro-koponen-2007}.

\color{DarkOrchid}\subsubsection{Decoupling Routing And Forwarding}\color{Black}

Although, the basic idea of using a binary system of Data and Interest messages for routing and forwarding data objects in the network has built a strong foundation for many ICN networks design, some prominent ICN proposals have also suggested architectures that aim to separate multicast group management, source discovery, and routing from the forwarding of Data messages in the network. This approach itself has appeared in three different ways which also have synergies with each other. 

\begin{itemize}[leftmargin=0.3cm]
    \item In the first way that is used by very prominent ICN architectures like SAIL \cite{sail-edwall-2011} and PSIRP/PURSUIT \cite{psirp-fotiou-2012, psirp_intro-tarkoma-2009, pursuit-trossen-2012}, a separate network of rendezvous nodes receive Interests (i.e., subscription requests) and manage multicast groups by allowing multiple recipients in the network to subscribe to the same data object name. At the same time, the rendezvous network also interact with a central or distributed topology management system to discover the network topology and perform routing. Once a subscription request arrives, the rendezvous network with the help of the topology manager can locate a matching data source/provider (i.e., publisher), find the forwarding path to the interested subscribers, and notify the publisher of the new request (if one is not already pending in the network). The rendezvous network also inform the data producer source of the calculated forwarding path which can be embedded in the Data messages header. All other forwarder nodes in the forwarding pathway can read the path information in the header and forward the messages accordingly \cite{psirp-fotiou-2012, psirp_intro-tarkoma-2009, pursuit-trossen-2012}. We note that this approach can natively support multicast forwarding of Data messages in the network. As we discussed in section 4.4., the PSIRP/PURSUIT architectures use Bloom Filters for encoding and compressing the forwarding paths which lead to a stateless forwarding plane design. 

    \item In the second way used in architectures like NetInf \cite{netinf-dann-2013} and Mobility First \cite{mobility_first-ray-2012}, a separate name resolution system (e.g. GNRS in Mobility First) is used to map publishers and unique data objects with network locators \cite{icn_locators-barak-2014}. In this approach, data recipients or the subscribers can query the same name resolution system to obtain the associated publishers locators for a given data object. Such information can then be used by the subscribers to send GET messages to the network that will be delivered to the given publishers. The publishers then send their data to the corresponding subscribers using the Reverse Path Forwarding method. We note that this approach can natively support multicasting as multiple subscribers can send requests for the same data object name to the network publishers.

    \item  In the third way, a software defined ICN controller is used that fully decouples the network control and data planes from each other. In this approach, data subscription information is shared with the ICN controller. If this is a pure ICN network, the controller can now use topology manager to perform routing and find associated data sources/providers in the network that match the subscription request (i.e., data object name) and inform any of them to provide its data. In this scenario, the source can get a forwarding path for stateless forwarding or the network controller can instruct the data switches. If this is a shared data plane environment where ICN and IP flows are handled at the same time, the ICN and IP network controllers are typically linked to handle the ICN requests. The latter has been also used in several proposals that are proprietary technologies.
    
\end{itemize}

%% file: content/goals.tex
\color{RedViolet} \section{Design Goals And Considerations} \color{Black}

\begin{figure}[h!]
  \includegraphics[height=5cm, width=6cm]{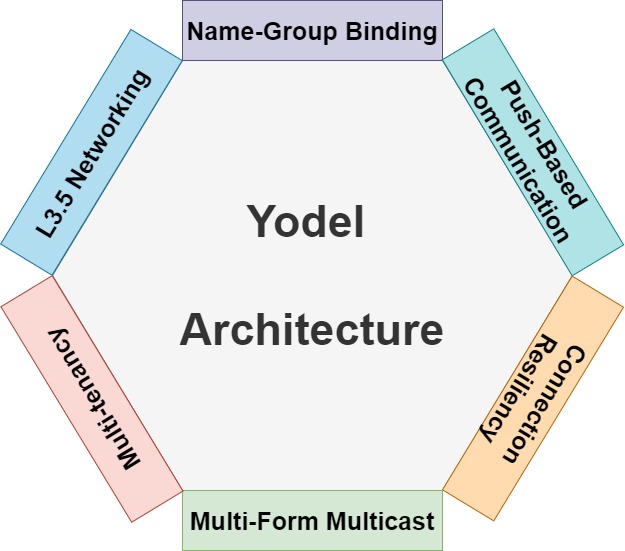}\bf\color{DarkOrchid}
  \centering
  \caption{\color{Black}\small Yodel Design Goals and Considerations}
\end{figure}

In section 2, we conducted a review of existing multicast group management methods in host-centric and information-centric networking architectures. Additionally, we explored the approaches taken by these architectures in dealing with multicast routing and forwarding. The objective was to gain a deeper insight into the design principles and decisions undertaken by these architectures to deliver network-level multicast services. In this section, our focus is on defining the design goals that shape the Yodel architecture and constitute its functional core \cite{yodel-morteza-2023}. These goals, as shown in Figure 1, follow our vision and motivations for building a name-based multicast network architecture that can meet the needs of current and future Internet applications. 

\color{DarkOrchid} \subsection{Name-Group Binding} \color{Black}

As stated in section 1, name-based networking is primarily proposed by ICN as an alternative method of identifying and routing data objects in the network. Information-centric networks usually adhere to the principle of name-data binding, indicating that each unique data object and all its identical copies in the network must be assigned a globally unique name identification \cite{rfc8793_icn_naming-wissingh-2020}. Communication is then about transmitting unique data objects in the network with senders and receivers (a.k.a. data producers and consumers) using any given name for communicating the associated data object. The network then utilizes the uniqueness of names for locating binding data objects (i.e., finding potential data producers), and for routing and delivery of those objects to their corresponding data consumers \cite{ICN_survey-boutaba-2012, ICN_survey-ahlgren-2012}.

As it can be inferred, the name-data binding principle can support multicast service abstractions, enabling multiple data consumers to concurrently receive the same data object by subscribing to its unique name \cite{ICN_survey-boutaba-2012, ICN_survey-ahlgren-2012}. A major concern, however, is the scalability of the routing tables as the so called name-data binding principle can cause large number of multicast groups to emerge, each revolving around the distribution of a distinct data object in the network \cite{rfc8793_icn_naming-wissingh-2020}. In other words, the number of multicast groups in the network is proportional to the number of distinct data objects. Of course using name aggregation \cite{rfc8793_icn_naming-wissingh-2020} supported by some ICN architectures \cite{icn_locators-barak-2014} may partially help reducing routing table sizes. Nonetheless, applications can still suffer from the scalability implications of the name-data binding principle. \bf{Yodel aims to address} \normalfont this challenge by modifying the ICN's name-data binding principle through the introduction of the name-group binding principle, indicating each multicast group in the network is assigned a globally unique name identification which also identifies all the group's data objects within the network, irrespective of their similarities and differences. In this approach, data producers and consumers can use any given name for joining an associated multicast group. The network then utilizes the uniqueness of multicast group names for identifying binding data objects, and for routing and delivery of those objects to the group's data consumers. Using name-data binding principle, the number of multicast groups can grow independently of the volume of data in the network.

In addition to addressing routing scalability, we believe that the purpose and semantic of name-group binding principle in Yodel can further help to simplify multicast applications logic, streamline designing complex multicast service abstractions, ease multicast group management, and facilitate routing of named data objects, specially in dynamic application environments, that are normally very difficult to achieve in the presence of name-data binding principle \cite{rfc8793_icn_naming-wissingh-2020}.

\color{DarkOrchid} \subsection{Multi-Tenancy} \color{Black}

\textbf{The Yodel architecture offers multi-tenancy}, enabling organizations and application developers to use logically separate networking environments that share a common Yodel multi-domain physical network infrastructure, possibly spanning over multiple network service providers. Multi-tenancy in Yodel can benefit organizations and application developers in various ways of: 1) facilitating network customization, 2) enabling name reusability, 3) lowering the cost of maintenance and network management, and 4) improving privacy. We note that without multi-tenancy, data from all applications and organizations typically traverse the same networking environment, leaving endpoints primarily responsible for preserving the privacy and integrity of communication (e.g., using end-to-end encryption). With multi-tenancy in place, logically separate networking environments can keep the data of independent applications and organizations isolated from each other which can consequently reduce the likelihood of compromising data and minimize the risks associated with communicating over a shared networking infrastructure. 

\color{DarkOrchid} \subsection{Multi-Form Multicast} \color{Black}

Multicast applications often require different multicast communication models for efficient distribution of dissimilar data types (e.g., images, audio, video, etc.) and for effectively addressing application-specific functional requirements. For instance, podcasts can effectively utilize a single-source (one-to-many) multicast communication model. Internet of things monitoring applications, on the other hand, typically favor a multi-source communication model, allowing all data sources to concurrently transmit their data to the same group of recipients. Other applications like video/audio conferencing may necessitate an even more flexible form of multicast communication, enabling mutual communication between all participants in an online conference or meeting room. \textbf{As a key design goal}\normalfont, the Yodel architecture can support multiple multicast service models, each enabling a different form of multicasting named data objects in a multicast group. Together, these services aim to support wide-ranging multicast applications needs. Further, Yodel can support simultaneous use of its distinct multicast services over the same Yodel-managed network infrastructure by allowing different tenants to use different multicast service models. We will later see that the Yodel software-defined approach of managing the network control and data planes (see section 4) as well as its hierarchical routing model (see section 5) play a fundamental role in reducing routing and infrastructural complexity that can be caused by concurrently supporting distinct multicast services over a common networking infrastructure.

Basically, each multicast service model typically specifies a different set of goals and metrics and may require different routing and group management protocols to function properly like the PIM-SSM \cite{ssm-bhat-2003} and PIM-SM \cite{pim_sm_motivation-deering-1996} used in IP Multicast to support single and multi-source multicast services. Some service models may even require additional infrastructure equipment to meet their specific multicast needs. Such increased heterogeneity of requirements and the sophistication involved in addressing them can complicate concurrent support of distinct multicast service models over a common networking infrastructure. Gradually, such complexities may even contribute to architectural ossification diminishing the utility of multicast services \cite{ossify-handley-2006, ossify-turner-2005, Internet-clark-2018}. We will discuss later in sections 4 and 5 that not only the Yodel architecture is capable of handling the heterogeneity of distinct multicast services but it is also extendable to support other forms of multicast service models without needing to change the Yodel architecture.

\color{DarkOrchid} \subsection{Push-Based Communication}\color{Black}

\textbf{Yodel provides push-based communication} by utilizing a Fire-and-Forget interaction model \cite{fire_n_forget-procedia-2020, fire_n_forget-voelter-2003}. In this method, data producers in a multicast group send their data to the network. The Yodel network then takes the responsibility of delivering the data to the group's data consumers without data producers needing to know the consumers nor receiving a feedback from them or the network. Similarly, data consumers in a multicast group do not know the data producers nor the existence of the data in the network. They also do not request the data. Instead, they receive the data once it becomes available to the network. Basically, communication is driven by a best-effort push-based asynchronous semantic \cite{fire_n_forget-procedia-2020, fire_n_forget-voelter-2003}. We believe that the Yodel's push-based approach can provide an effective name-based communication method for addressing the multicast needs of modern Internet applications, especially in dynamic application environments. In comparison, existing ICN architectures use consumer-driven communication model. The consumer-driven model is a pull-based communication method where data consumers must send a request every time they need a data object to retrieve the object from the network. This method works very well in situations where a data object exists in the network \cite{push_based-nour-2019}. 

The problem, however, arises if the data object becomes available in the future or does not exist at the time of the request (very common in dynamic application environments). In these situations, routing algorithms usually fail to locate the data and require data consumers to actively sending data requests (a.k.a., Interests) to ensure the delivery of the data object once it becomes available. In compare, the Yodel push-based communication method does not require data consumers to request the data as stated before. Instead, they receive the data once it becomes available to the network. This way routing only happens when the data is available in the network without overloading the network with unnecessary request messages. Another important aspect of the Yodel push-based method is the anonymized nature of communication between data producers and consumers in the multicast groups. Most Internet multicast applications (e.g., audio/video conferencing, IoT, social media, etc.) work by dynamically generating data that will be consumed by some interested data consumers regardless of who is the data producer. In other words, the data (i.e., WHAT) is more important than the actual provider of the data (i.e., WHO). For these applications Yodel can provide an ideal communication model. 

There might be also existing multicast applications that rely on recognizing the provenance of data for building trust and identification purposes. These applications can also benefit from Yodel communication model as source verification can be achieved through complementary methods like key-based encryption and verification \cite{key_based_icn-thamb-2012,key_based_icn-tschudin-2018}. Further, multicast group membership in Yodel requires authentication (see section 5) which can ensure data producers and consumers are pre-authorized to communicate. Membership authentication minimizes the associated security risk factors of communicating anonymously for all multicast applications, especially those with tighter security concerns, while benefiting the applications with advantages of push-based communication method.

\color{DarkOrchid}\subsection{Connection Resiliency}\color{Black}

\textbf{A key Yodel architectural goal} is to keep users programs (i.e., data producers/consumers) connected with network services and data in situations where they become temporarily disconnected or have difficulty remaining connected. End-user devices, called Hosts, are where the users programs typically run in Yodel. These programs usually share the Host's network interface and connectivity to access the network's services and data. Therefore, Host disconnectivity can affect all user programs running on a Host. Yodel mitigates this problem by using an in-network digital twin of a Host, called the Host Twin to keep the Host's programs connected in the event of Host disconnection. A Host Twin can function as a Host in the moments of disconnectivity, keeping the Host's data consumers subscriptions and multicast group memberships active so that they can continue receiving data while the Host is offline. These data can be accessed once the Host's connectivity resumes. Of course data producers cannot continue providing data while the Host is offline but the Host Twin can keep data producers registrations and multicast group memberships active, so that data producers can seamlessly continue providing data once the Host connectivity resumes. 

\color{DarkOrchid}\subsection{Layer 3.5 Networking}\color{Black}

One of the notable challenges of name-based networking is its incapacity to seamlessly span across multiple domains without a clean-slate reconstruction of the existing internetworking infrastructures. The issue stems from a fundamental incompatibility between addressing and routing in name-based networking architectures and the current Internet architecture's host-centric addressing and routing mechanisms that makes name-based networking incapable of sharing the existing internetworking infrastructures with IP. Meanwhile Internet applications require multi-domain network support to meet their multicast needs, prompting the need for a practically viable solution. The complete reconstruction of the existing internetworking infrastructures is not ideal as it is costly, does not follow the current Internet deployment pathway, and lacks backwards compatibility, thereby complicating the adaption of name-based networking by existing Internet applications and systems \cite{Internet-clark-2018}. To avoid unnecessary changes in the existing internetworking infrastructures, several approaches like application-layer overlays \cite{overlay_icn-shail-2015}, hybrid routing infrastructures \cite{hicn-caro-2019}, and name-based BGP extensions \cite{N_BGP-aldaoud-2022} have been proposed so far by the ICN community. Despite the level of flexibility and adaptability that each approach can provide, none have resulted in a large-scale deployment of name-based networking so far, leaving the doors open for more innovations.

\begin{figure*}[h]
  \includegraphics[height=5cm, width=16cm]{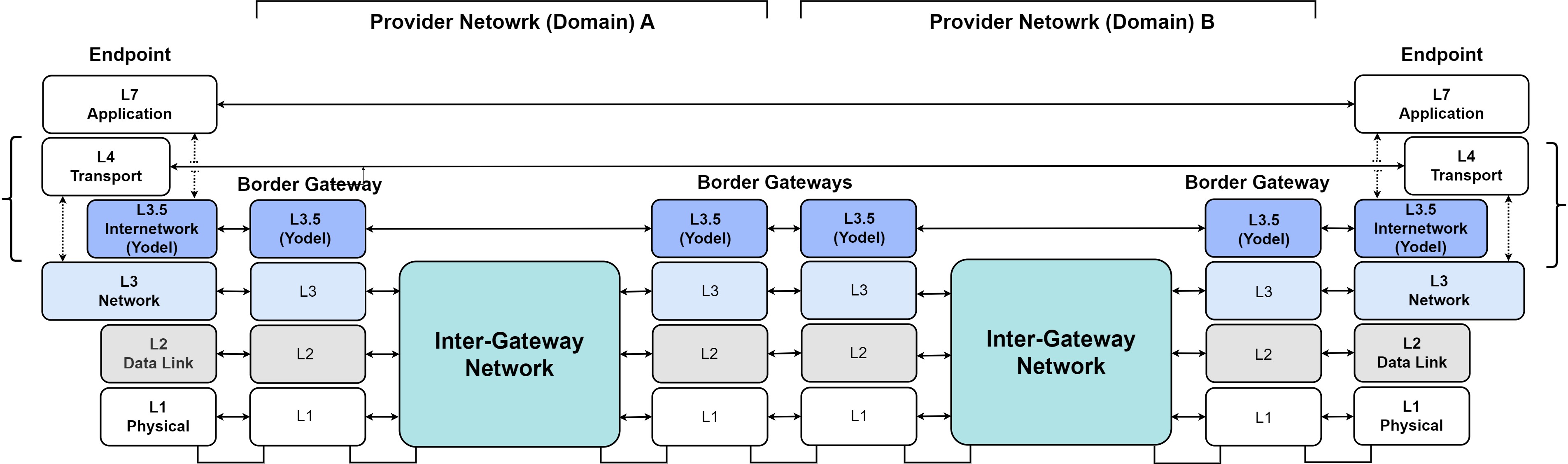}\bf\color{DarkOrchid}
  \centering
  \caption{\color{Black}\small The current Internet protocol stack, demonstrating the adaption of a new layer 3.5 for handling interdomain networking.}
\end{figure*}

In a distinct and Internet-friendly approach, \textbf{Yodel leverages layer 3.5} for achieving name-based multi-domain networking \cite{L3.5-shenker-2019}. Using separate layers for handling different physical media (data link), networks (intradomain), and network of networks (interdomain) with each layer having multiplicity of coexisting protocols and self-defined addressing methods was first discussed by INWG in 1976 \cite{inwg-davies-2010, inwg-day-2011, inwg-russell-2013} but has not been practically used until 2008. In 2008, the idea has been successfully adapted by Recursive InterNetwork Architecture (RINA) \cite{rina-grasa-2011, rina-john-2007} and has recently been revisited and redefined by James McCauley et. al, in their seminal paper proposing a new internetworking layer (i.e., L3.5) for the current Internet protocol stack \cite{L3.5-shenker-2019}. Basically, the proposed layer 3.5 is defined to be an independent internetworking layer, running above layer 3 (see Figure 2), with self-defined methods of addressing, routing, and forwarding data across separate domains. However, within each domain, a L3.5 network protocol like Yodel can directly use the services of any intradomain routing protocol (e.g., IP, IP Multicast) to move data inside the domain. This way, a layer 3.5 network architecture can reuse the existing networking infrastructures provided by the layers below (e.g., IP-based network infrastructure) while steering the traffic across interconnected domains.

In layer 3.5 networking approach, layer 3.5 messages/packets can be encapsulated inside layer 3 PDUs and the layer 3.5 header must be processed by the network endpoints for accessing L3.5 data and networking services, and by the border gateways in each domain for steering the traffic across separate domains, as shown in Figure 2. We note that in a L3.5-capable networking environment, network applications have the option to either bypass layer 4 transport protocols and directly use the services of a L3.5 network protocol \cite{inwg-day-2011, L3.5-shenker-2019} for end-to-end communication (the case for Yodel), or indirectly using layer 3.5 networking services through any L4 protocol that is compatible (see Figure 2). Applications can also choose a specific L3.5 protocol, among coexisting choices to meet their end-to-end communication needs.

We believe, designing Yodel as a L3.5 network architecture presents an opportunity to build a multi-domain multicast infrastructure for the current and future Internet applications without dependency on IP and BGP for internetworking, and with multiple multicast service models, as presented in section 3.3. Since layer 3.5 operates below the transport layer, it also does not have the overloading features presented by application-layer overlays. L3.5 networking also makes Yodel compatible with the current Internet deployment pathway (i.e., no clean-slate reconstruction of the Internet infrastructure), simplifies implementation of multicast services, and allows Yodel to rapidly integrate separate network domains regardless of their type and organizations. The latter also allows Yodel to grow independently of the scale of the current Internet architecture as it can use any layer 3 protocol (e.g., protocol X) for intradomain networking. In other words, Yodel can remain as a general multicast service option for the Internet applications as the Internet infrastructure evolves.

%% file: content/architecture.tex
\color{RedViolet} \section{Yodel Architecture}\color{Black}

In section 3, we described the Yodel design goals and explained the role they play in the Yodel architecture and the logic behind them. In this section, we explain how we meet these goals through a set of design decisions that we made to realize them.

\color{DarkOrchid}\subsection{How Yodel Resolves Architectural Challenges}\color{Black}

The Yodel architecture, as defined by its design goals, encompasses various technologies for addressing architectural challenges ranging from managing multi-tenancy and concurrent use of diverse multicast service models to implementing name-group binding and layer 3.5 networking. Recognizing the need for effective problem-solving methods amidst such architectural complexity, we have made two fundamental decisions to establish a robust foundation for addressing architectural challenges in Yodel \cite{yodel-morteza-2023}.

First, we leveraged Software-defined Networking (SDN) as a powerful framework for dynamic management of the entire Yodel multi-domain network infrastructure and services \cite{sdn-goran-2016,sdn-nick-2009}. Second, we designed and used a conceptual model that logically organizes the Yodel network infrastructure in Valleys, Namespaces, and Communities, as shown in Figure 3. SDN stands out as an excellent approach for managing dynamic networks like Yodel with complex network services \cite{sdn-goran-2016} and the Yodel conceptual model provides a cohesive architectural organization intended to empower Yodel with divide and conquer problem-solving capabilities \cite{divide_conquer-baase-2009}. Divide and conquering is an efficient method for handling large-scale and complex architectural challenges (e.g., naming, routing, caching, and security) by breaking them into smaller pieces/scales to solve \cite{divide_conquer-baase-2009}. It is particularly useful in software-defined networks as the logically centralized networking view used for network management cannot always scale well to large and realistic networking environments \cite{divide_conquer-schneider-2021}. In this sense, the Yodel conceptual model's approach of enabling divide and conquering can improve the efficiency, effectiveness, and performance of Yodel's software-defined network management, especially in large-scale networking environments.

\begin{figure*}[h]
  \includegraphics[height=9.75cm, width=16.75cm]{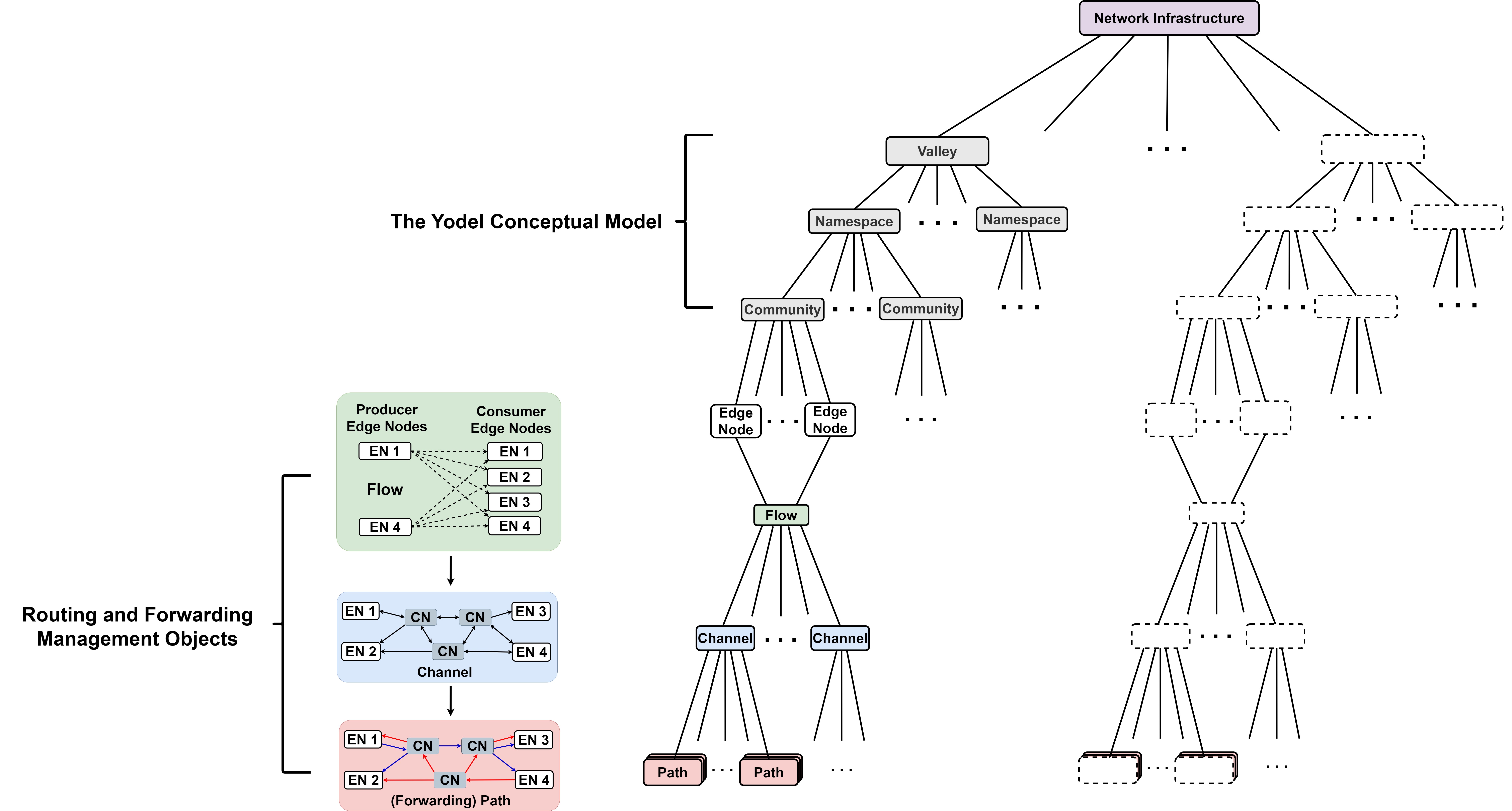}\bf\color{DarkOrchid}
  \centering
  \caption{\color{Black}\small The Yodel management tree, featuring the Yodel conceptual model as well as the routing and forwarding management objects.}
\end{figure*}

\color{DarkOrchid}\subsection{The Yodel Conceptual Model}\color{Black}

As stated before, Valleys, Namespaces, and Communities are the three main components of the Yodel conceptual model. As shown in Figure 3, these logical components are vertically related to each other. In the following, we progressively describe these components, their relations, and the roles they play in enabling divide and conquering in Yodel from the most fundamental component (i.e., Community) upwards.

\color{DarkOrchid}\subsubsection{Community}\color{Black}
A Community is a multicast group of data producers and consumers that share an interest in communicating named data objects. Communities have specific attributes. A Community has a name that identifies and distinguishes it from other Communities within its parent Namespace. A Community name is immutable and is also used to identify and route the Community's named data objects and subsequently all the data messages that carry those data objects in the network. That being said, all data objects that are communicated within a Community have the same name identification (i.e., the Community name) but they may have different data (i.e., realizing name-group binding principle). A Community also has a multicast service model (i.e., communication mode) which is defined by the multicast service model of its parent Namespace. In other words, all Communities within a Namespace have the same communication mode.

\color{DarkOrchid}\subsubsection{Namespace}\color{Black}
By design, every Community belongs to a Namespace, as shown in Figure 3. In other words, a Namespace is a Community of (related) Communities. These Communities have similar communication requirements (i.e., privacy needs, multicast service model, and caching) and so their needs can be met by the same set of customized network environments (i.e., Namespaces). For the choice of multicast service model, a Namespace can use a specific multicast service model from the set of network multicast services supported by its parent Valley. To enforce privacy, a Namespace implements a set of privacy rules, called Visibility Terms, which govern the access of users of the parent Valley to the Namespace. An Open Namespace can be accessed by all users of the parent's Valley whereas a Protected Namespace can be only accessed by the users of the parent's Valley who are authorized by the Namespace admin user to access the Namespace. A Namespace Visibility Term can be changed during the life cycle of the Namespace. Any user who can access a Namespace can join its existing Communities or create new ones. A Namespace carries an admin-defined immutable name that sets it apart from other Namespaces within its parent Valley.

\color{DarkOrchid}\subsubsection{Valley}\color{Black}
A Valley is a collection of Namespaces that is served by a single dedicated Yodel network. The Yodel network corresponding to a Valley is an independent, logical, virtual, and private environment with its customized control plane, dedicated data plane resources, associated policies, and supported multicast service models that can be selected from the set of multicast service models provided by the Yodel architecture. As shown in Figure 3, all Valleys share an overall Yodel physical network infrastructure. Within a Yodel network infrastructure, Valleys are identified by unique and immutable names that distinguish them from coexisting Valleys. Each Valley also consists of several users who can access the Valley's Namespaces (i.e., based on the Namespaces Visibility Terms) or can create new ones. Creating a Namespace makes the user the Namespace admin. We note that users of a Valley are by default users of the Yodel network infrastructure that are permitted access to the Valley by its admin user. An infrastructure user who creates a Valley is the admin user for that Valley. 

\newpage

\color{DarkOrchid}\subsubsection{Divide and Conquering In Yodel Conceptual Model}\color{Black}

As it can be inferred, Valleys, Namespaces, and Communities each play a key role in enabling divide and conquering and simplifying network management in Yodel. Valleys, for instance, make the very large and centralized view of the network easier to manage by dividing the global view of the network infrastructure and multicast groups into logically separate and smaller views to manage. Namespaces further break down the management views built by Valleys to smaller scales for handling security, privacy, and finer multicast group requirements (e.g., multicast service model). In this regards, Namespaces act similar to the notion of Control Groups (Cgroups) and Namespaces in the context of Linux operating systems \cite{os-chak-2023} but are technically different entities. Communities even go further by making it easier to meet more specific multicast group challenges such as routing and addressing quality of service requirements. We will see later (in section 5) that abstracting multicast groups as Communities has also empowered Yodel to extend its divide and conquering approach to simplify the management of routing and forwarding functions in the network, using Flow, Channel, and Path objects (see Figure 3).

\color{DarkOrchid}\subsubsection{Addressing Multi-tenancy}\color{Black}

In addition to enabling divide and conquering, the combination of Valleys and Namespaces can also enable multi-tenancy (see section 3.2), where Valleys allow independent virtual networks to coexist in a shared networking environment, and Namespaces provide name reusability and offer organization, customization, and protection for multicast groups (i.e., Communities). We note that name reusability is an intended outcome of enabling multi-tenancy (see section 3.2) in Yodel and is also a very important feature for a name-based network architecture as names must be always unique in the network. 

In a single tenant network, each name can be only used once to create a multicast group. However, in a multi-tenant network like Yodel, a multicast group name must be only unique in individual Namespaces. So, names can be reused in other Valleys and Namespaces. Such capability allows organizations and application developers to freely use names in their programs without needing to worry about the challenges of name uniqueness. It is also noteworthy to mention that the name reusability offered by Yodel Valleys and Namespaces is fundamentally different from content scoping proposed by some ICN architectures \cite{psirp-fotiou-2012} as: 1) identical names in separate Yodel Valleys and Namespaces do not necessarily identify the existence of the same data objects in the network nor they are at all related to each other, and 2) Yodel does not recognize name-data binding principle and consequently does not understand the name and content uniqueness relationship.

\color{DarkOrchid}\subsection{The Yodel Software-Defined Network Infrastructure}\color{Black}

As stated in section 4.1, Yodel leverages software-defined networking to manage the entire Yodel multi-domain network infrastructure. That being said, the Yodel architecture features separate control and data planes as specified in the followings.

\color{DarkOrchid}\subsubsection{The Yodel Control Plane}\color{Black}

The Yodel architecture features a logically centralized network controller that manages all infrastructure resources belonging to different domains and is responsible for handling the compute and networking resources used to support one or more isolated Valleys (see section 4.2.3).  

\begin{figure}[h!]
  \includegraphics[height=4.5cm,width=7.75cm]{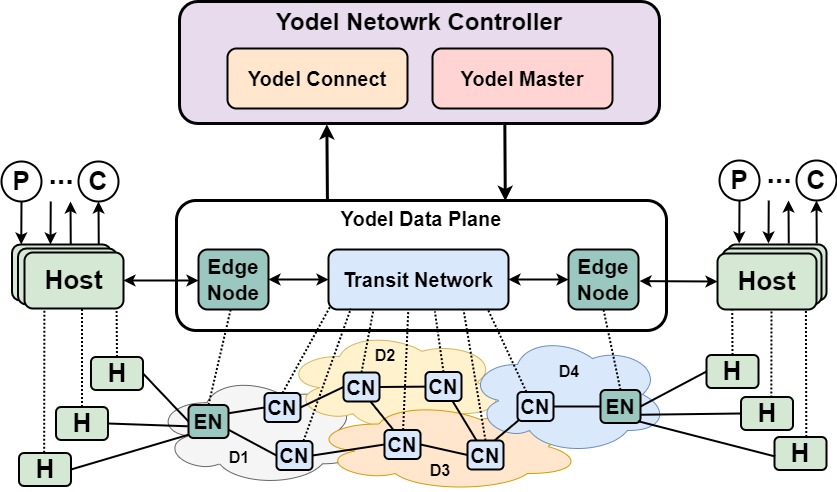}\bf\color{DarkOrchid}
  \centering
  \caption{\color{Black}\small The Yodel Network Infrastructure (H=Host, EN=Edge Node, CN=Connector Node, and P/C=Data Producer/Consumer).}
\end{figure}

The Yodel network controller is composed of two subcontrollers, the Yodel Master and the Yodel Connect, with each play a different role and handle a different set of architectural responsibilities. The Yodel Connect is responsible for: 1) configuring Hosts to connect to the Yodel Edge nodes, 2) validating users access to Valleys and Namespaces (i.e., user management), and 3) maintaining a graph topology of Yodel Edge and Connector Nodes across the Yodel multi-domain network infrastructure to assist in routing as we will discuss later in section 5. We note that user management in Yodel ensures only authorized users can access the network infrastructure, Valleys, and their Namespaces and services. The Yodel Master manages Valleys and their associated Namespaces and Communities as well as routing of named data objects within individual Communities with respect to their multicast service model (see section 4.4). We note that individual Valleys and their services and Namespaces are managed independently from each other. This arrangement is made to allow Valleys to have their own customized control planes. We also note that the two Yodel subcontrollers interact with each other using a controller-specific message queuing bus.  

\color{DarkOrchid}\subsubsection{The Yodel Data Plane}\color{Black}

The Yodel data plane consists of two different nodes. Edge Nodes of individual domains that connect data producers and consumers running on end-user devices (i.e., Hosts) to Valleys and their Namespaces and services. Connector Nodes that interconnect Edge Nodes across separate domains. We refer to the network of these Connector Nodes as the Yodel Transit Network (see Figure 4). Individual infrastructure domains can also use Connector Nodes to become a transit domain like D2 and D3, as shown in Figure 4. Both Edge and Connector Nodes have a controller agent that interacts with the network controller for programmability and configuration and they maintain a Forwarding Information Base (FIB) with several forwarding tables that enable forwarding of named data objects in Yodel Communities. We will learn more about these tables and their roles in each Yodel node in section 5. Edge and Connector Nodes also form a graph that provides the connectivity needed by the Valleys. This graph is maintained by Yodel Connect as stated in section 4.3.1. We note that individual infrastructure domains can use as many Edge Nodes as they want to take Yodel Valleys and services to users in different geographical locations. Similarly, such domains can also use as many Connector Nodes as needed to interconnect with other domains.

\color{DarkOrchid}\subsubsection{Infrastructure Setup}\color{Black}

All Hosts, and every Yodel Edge and Connector Nodes must go through a multistage bootstrapping process before they join the network. Initially, each node must self generates a unique Yodel Node ID (YNI). A YNI is a 10-byte Pseudo MAC address that serves as a layer 3.5 address for a Yodel node in the network. The YNI is basically composed of the node MAC address and a time stamp, as shown in Figure 5.

\begin{figure}[h!]
  \includegraphics[height=1.15cm,width=7.5cm]{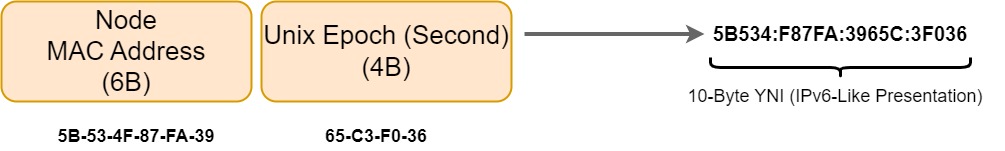}\bf\color{DarkOrchid}
  \centering
  \caption{\color{Black}\small The Yodel YNI in its default IPv6-like presentation.}
\end{figure}

The MAC address is 6 bytes and can be the MAC address of any of the node's network interfaces or it can be a randomly generated MAC address produced by the node. The time is 4 bytes and is the Unix time calculated in seconds from 12am UTC on January 1st, 1970. The two combine make the node's 10-byte unique YNI which we show in an IPv6-like format by default (see Figure 5). We note that address uniqueness is achieved in Yodel YNIs as MAC addresses are unique and in case of finding duplicate addresses (e.g., by error) the time component can ensure that addresses can remain unique. We also note that Yodel layer 3.5 addresses (YNIs) are intended to be agnostic of the type and the number of network interfaces a Yodel node has to simplify routing and enable a process we refer to as Strategic Forwarding (described in section 5.4.2). Moreover, YNIs also decouple L3.5 addresses in Yodel from the changes of addresses in layers below (e.g., IP address).

\begin{figure}[h!]
  \includegraphics[height=3.5cm,width=7.75cm]{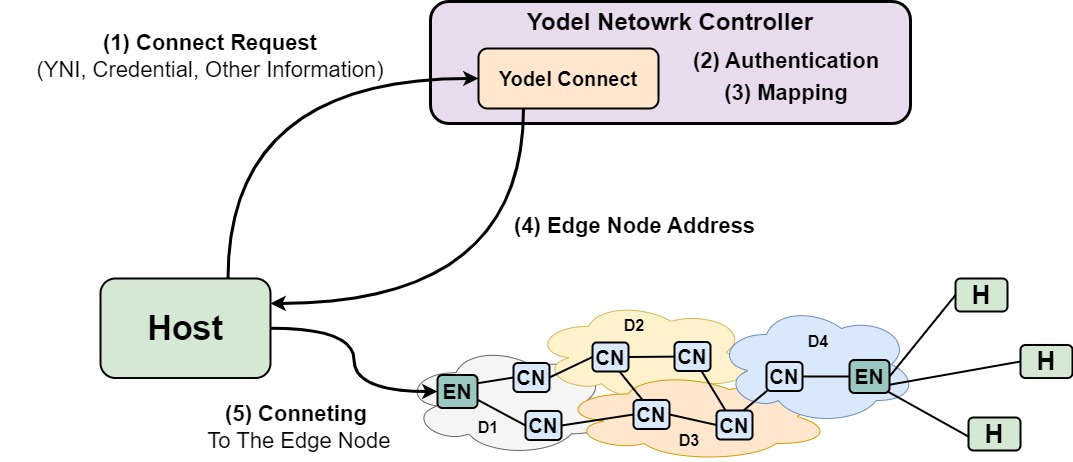}\bf\color{DarkOrchid}
  \centering
  \caption{\color{Black}\small Demonstrating the Host Provisioning Process (H=Host, EN=Edge Node, and CN=Connector Node).}
\end{figure}

Next, Hosts contact the Yodel Connect to find the (best) network access point (i.e., Yodel Edge Node) to connect to the Yodel network. This process, as shown in Figure 6, is known in Yodel as Host Provisioning. In response, the Yodel Connect performs two tasks. First, it verifies the Host's admin user credential and second, if verification succeeds, The Yodel Connect maps the Host to a Yodel Edge Node in the network. We note that \say{best} here refers to the most ideal Yodel Edge Node that can satisfy connectivity constraints such as vicinity, Edge Node capacity, latency, etc. The aim is to allow Hosts to describe their preferences in an Intent-driven approach. Finally, in the last stage of bootstrapping process, the Host receives the Yodel Edge Node address from the controller and contacts the given Yodel Edge Node to connect. At this stage, the Yodel Edge Node can also ensure that the Host YNI is unique among the connected Hosts at the same Edge Node. For reasons that we learn later in section 5, a Host YNI can be unique within each Yodel Edge Node. In case an address duplication happens, the Yodel Edge Node requests the Host to regenerate its YNI. 

Much like Hosts, Yodel Edge and Connector Nodes also interact with Yodel Connect for bootstrapping and to share their YNI and those of their neighbors with the network controller (i.e., Yodel Connect). If any changes occur in the neighbors, the node must inform the controller of the changes. In addition, Yodel Edge and Connector Nodes provide other pieces of information about themselves such as their available bandwidth, computing power, and storage capacity which they can also regularly update. The controller uses this information to first create and maintain the graph of Edge and Connector Nodes (see section 4.3.1), and second to maps Hosts with Edge Nodes. Yodel Edge and Connector Nodes can also connect to each other in two ways. First, by manually providing an endpoint address to connect, and second, through automatic redirection via the Yodel Connect Node in a process similar to the Host provisioning \cite{yodel-morteza-2023}.

\color{DarkOrchid}\subsubsection{Adapting Layer 3.5 Networking}\color{Black}

As described in section 3.6, the Yodel architecture is designed to work in layer 3.5. In this approach, L3.5 infrastructure devices coordinate the movement of data in between separate infrastructure domains while using the services of the layers below for moving data inside each domain, that is in between L3.5 devices inside a domain (see Figures 2 and 4). To achieve this goal, we need to introduce minimal software changes in the current networking stack of the devices that process the L3.5 semantic (i.e., Hosts, Edge, and Connector Nodes in Yodel). 

\begin{figure}[h!]
 \includegraphics[height=4.75cm,width=5cm]{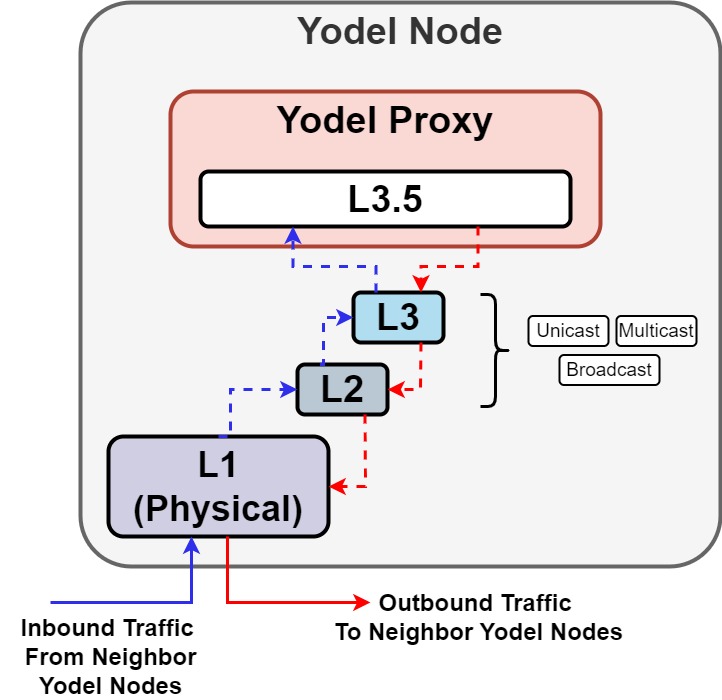}\bf\color{DarkOrchid}
  \centering
   \caption{\color{Black}\small A high-level view of the Yodel Proxy}
\end{figure}

These changes require L3.5 devices to fulfill two key objectives alongside other designated architectural responsibilities. First, they must be able to process the layer 3.5 headers which are encapsulated within the PDUs of the substrate layer 3 (see Figure 7) protocols, and second, these devices must have the ability to discover and use the networking services provided by the layers below to move data to their neighbor L3.5 devices inside a domain. For instance, the Edge Node in Figure 4 (D1) can forward its data messages to its two neighbor Connector Nodes using two separate unicast communications or a single-source IP multicast communication depending on the service availability in domain D1. We note that neighbor nodes can exchange their ability to use different networking services after being connected to each other as described in section 4.3.3. As we will explain later, Yodel has the privilege and ability to use different intradomain networking services, upon availability, to move data between two Yodel infrastructure nodes. 

To let the Yodel infrastructure nodes to fulfill the requirements set by the two aforementioned objectives, each node uses a component, called the Yodel Proxy. The Yodel Proxy, as shown in Figure 7, maintains the forwarding tables (i.e., FIB), manages the functions that process the Yodel layer 3.5 headers, and enable each infrastructure node to make an informed decision on how to move data messages to its neighbor nodes through Strategic Forwarding process that we will describe in section 5.4.2. On a Host, the Yodel Proxy also links data producers and consumers to Yodel Client Daemon (YCD). YCD is a client software that runs in the user space of a Host's operating system and handles the Host's networking and connectivity with the Yodel network. We will discuss the organization and functional components of the Yodel Proxy in more details in section 5.3. 

\color{DarkOrchid}\subsection{Yodel Multicast Services}\color{Black}

By design (see section 3.3), Yodel can support multiple multicast service models. These services are designed and can be extended to enable various forms of multicast interactions between data producers and consumers in a given Community (i.e., multicast group). In this section, we briefly review the multicast services that the Yodel architecture can currently provide. Later in appendix A, we revisit these services to provide a more detailed discussion of the their specifications and relevant use cases. Appendix A also includes a table summary of all the services.

\color{DarkOrchid}\subsubsection{Yodel Single Source Multicast (Yodel-SSM)}\color{Black}

The Yodel-SSM is a Yodel multicast service model that enables a single data producer in a Community to send its name data objects to all data consumers in that Community, as shown in Figure 8. The Yodel-SSM service uses one active data producer for a Community while keeping other available data producers in the Community (if any) on-hold for use in situations where the current Community's active data producer becomes inactive or unavailable. The Yodel-SSM service is the most basic of all Yodel multicast service types.

\begin{figure}[h!]
  \centering
  \animategraphics[autoplay,loop,width=7.5cm]{1}{images/ssm/ssm-}{0}{8}\bf\color{DarkOrchid}
  \caption{\color{Black}\small Multicasting using the Yodel-SSM service. This diagram is animated. (P=data producer, C=data consumer, H=Host, EN=Edge Node, and CN=Connector Node).}
\end{figure}

\color{DarkOrchid}\subsubsection{Yodel Anycast (Yodel-AC)}\color{Black}

The Yodel-AC is a Yodel multicast service model that advances the Yodel-SSM basic service model by controlling which data consumers can receive the Community's data objects. Using the Yodel-AC service, a subset of all data consumers in a Community can receive the data objects produced by the Community's data producer, as shown in Figure 9. The service may also deliver each data object to a different subset of the Community's data consumers as the Yodel-AC service allows users to indicate a randomized selection or a dedicated subset of all data consumers as the recipients.

\begin{figure}[h!]
  \centering
  \animategraphics[autoplay,loop,width=7.5cm]{1}{images/ac/ac-}{0}{8}\bf\color{DarkOrchid}
  \caption{\color{Black}\small Multicasting using the Yodel-AC service. This diagram is animated. (P=data producer, C=data consumer, H=Host, EN=Edge Node, and CN=Connector Node).}
\end{figure}

\color{DarkOrchid}\subsubsection{Yodel Selective Source Multicast (Yodel-SLSM)}\color{Black}

The Yodel-SLSM is a Yodel multicast service model that advances the Yodel-SSM service model by introducing the notion of partitioning. Basically, the service partitions a Community into disjoint groups, wherein a single data producer in each partition can send its named data objects to all data consumers in that partition, as shown in Figure 10. We note that a partition may have more than one data producer but similar to Yodel-SSM, there is always one active data producer in the partition while the rest remain on-hold.

\begin{figure}[h!]
  \centering
  \animategraphics[autoplay,loop,width=7.5cm]{1}{images/slsm/slsm-}{0}{8}\bf\color{DarkOrchid}
  \caption{\color{Black}\small Multicasting using the Yodel-SLSM service. This diagram is animated. (P=data producer, C=data consumer, H=Host, EN=Edge Node, and CN=Connector Node).}
\end{figure}

The fundamental assumption in Yodel-SLSM service is that all data producers in a Community generate the same data objects. Thus, partitioning can reduce communication latency and distribute the load on the Community's data producers. We note that the Yodel-SLSM service requires the presence of at least two data producers in a Community to begin partitioning. The Yodel-SLSM service can be also seen as a special form of any source multicasting where more than one data producer in a Community can be selected to simultaneously provide their data. The Yodel-SLSM service can be also combined with the Yodel-AC service (a.k.a., Yodel-SLAC) to enable delivery of multicast data to a subset of data consumers in each partition. 


\color{DarkOrchid}\subsubsection{Yodel Multi Source Multicast (Yodel-MSM)}\color{Black}

The Yodel-MSM is a Yodel multicast service model that enables all data producers in a Community to send their named data objects to all data consumers in that Community, as shown in Figure 11. The assumption here is that different data producers in a Community can produce different data objects and the Community's data consumers need all instances of the Community's data regardless of their provenance, similarities, and the differences. 

\begin{figure}[h!]
  \centering
  \animategraphics[autoplay,loop,width=7.5cm]{1}{images/msm/msm-}{0}{8}\bf\color{DarkOrchid}
  \caption{\color{Black}\small Multicasting using the Yodel-MSM service. This diagram is animated. (P=data producer, C=data consumer, H=Host, EN=Edge Node, and CN=Connector Node).}
\end{figure}

The Yodel-MSM service can be also seen as a special form of any source multicasting where all data producers in a Community can simultaneously provide their data to the Community's data consumers. This form of communication is also common across application-layer messaging protocols like MQTT \cite{mqtt_rfc-oasis-2020} and AMQP \cite{amqp-vinoski-2006} for distributing telemetry and machine-to-machine communication. The Yodel-MSM service can be also combined with the Yodel-AC service (a.k.a., Yodel-MSAC) to enable delivery of multicast data to a subset of data consumers in a Community.

\color{DarkOrchid}\subsubsection{Yodel Many-to-Many Multicast (Yodel-MMM)}\color{Black}

\begin{figure}[h!]
  \centering
  \animategraphics[autoplay,loop,width=7.5cm]{1}{images/3m/3m-}{0}{8}\bf\color{DarkOrchid}
  \caption{\color{Black}\small Multicasting using the Yodel-MMM service. This diagram is animated. (M=data producer and consumer, H=Host, EN=Edge Node, and CN=Connector Node).}
\end{figure}

The Yodel-MMM, also called the Yodel-3M, is a Yodel multicast service model that enables every member in a Community to send its named data to all other members in that Community. In other words, every member can produce and consume data simultaneously, as shown in Figure 12. A member is a process (i.e., user program) running on a Host that can produce and consume data simultaneously. We use the word member to differentiate between a data producer/consumer process with the singular role of data production/consumption and a process that has the ability to perform both.

%% file: content/rf.tex
\color{RedViolet}\section{Data Delivery In Yodel}\color{Black}

In the previous section, we introduced the Yodel management framework, the Yodel control and data plane components and functions, and multicast services. In this section, we explain how these components work in concern to distribute named data objects in Yodel multicast groups (i.e., Communities).

\color{DarkOrchid}\subsection{Joining A Community}\color{Black}

The end-to-end communication in a Community begins with data producers and consumers to first join the Community (i.e., intended multicast group). The joining process is the same for all Communities but individual Community's multicast service model can specify distinct rules that control the number of data producers and consumers who can join the group (see section 4.4 and appendix A). In this section we simplify describing the Joining process by considering a general service model where multiple data producers and consumers can freely join a Community. We will later describe particulars of individual multicast services in appendix A.

\color{DarkOrchid}\subsubsection{Signaling A Yodel Edge Node}\color{Black}

Generally, each data producer/consumer has an application ID that obtains it from the Host YCD where it is linked. As we stated before, a data producer/consumer is a process running in the user space of a Host's operating system. To Join a Community, the data producer/consumer must send a request to the Host YCD specifying its application ID (AID), the given Community name, and the Namespace and Valley name/ID where the Community belongs to, as shown in Figure 13. We assume that the data producer/consumer has already been authorized to access the intended Valley and Namespace in a process described in details in \cite{yodel-morteza-2023} and therefore, has access to the intended Valley and Namespace IDs. For simplicity, we refer to a Valley and Namespace by using the abbreviation \say{VNS} in the future. Once a Host receives a Join request from a data producer/consumer, the Host records the data producer/consumer application ID for the given Community in the intended VNS along with the Valley ID of the request in the Producer Registration Table (PRT) for data producers and in Consumer Registration Table (CRT) for data consumers. These two tables are the Host's primary source of knowledge about its applications registration status with different Communities and are an important part of the Host's Forwarding Information Base, as shown in Figure 13. After processing the request, the Host then forwards the Join request to the Yodel Edge Node where it is connected for additional processing by adding its YNI to the request. 

Upon receiving the Host's message, the Yodel Edge Node that received the message performs two main tasks. First, it checks whether it has previous records of the same or other Hosts in its reach (i.e., connected to the same Edge Node) requesting for data production (if the request comes from a data producer) or consumption (if the request comes from a data consumer) in the same Community and intended VNS. Each Yodel Edge Node in the network keeps records of Hosts with data producers (i.e., producer Hosts) in Pending Producer Table (PPT) and Hosts with data consumers (i.e., consumer Hosts) in Pending Consumer Table (PCT) along with the corresponding Namespaces and Communities of their registration in separate FIB instances that pertain to separate Valleys (i.e., Valleys have independent data plane resources as described in section 4.2.3). These tables are also an integral part of the Yodel Edge Node's Forwarding Information Base, as shown in Figure 13. 

\begin{figure*}[h]
  \includegraphics[height=14.25cm, width=16cm]{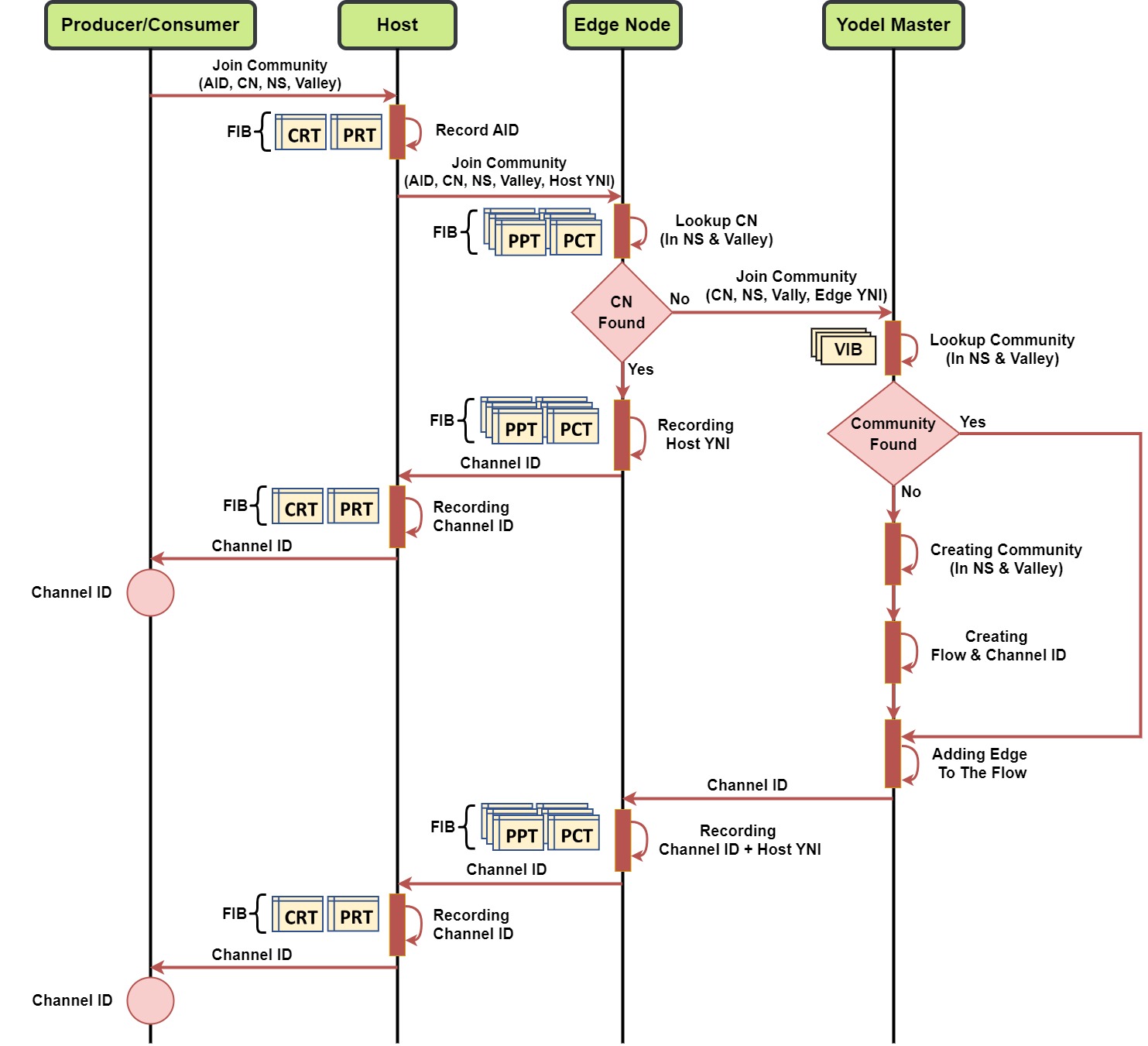}\bf\color{DarkOrchid}
  \centering
  \caption{\color{Black}\small Sequence diagram showing data producers and consumers joining a given Community. (AID=Application ID, CN=Community Name, NS=Namespace Name/ID, and VIB=Valley Information Base).}
\end{figure*}

Second, if the Edge Node finds prior records of data production or consumption (depending on the type of the request) for the same Community in the intended VNS, it adds the Host YNI to its current records for the given Community and Namespace in the FIB of the intended Valley's PRT or CRT table, depending on whether the request is for a data production or consumption respectively, and responds to the Host by providing a Community's associated Channel ID, as shown in Figure 13. We will learn more about the Channel IDs and their usage in routing and forwarding in the future. In response, the Host also records the Channel ID for the requesting data producer or consumer application in the corresponding PRT or CRT table depending on the type of the request (i.e., data production/consumption), and informs the data producer/consumer of the ID of the Community's Channel, as shown in Figure 13. The Channel ID will be used in the future by data producers to send their data to the given Community and by data consumer to receive the Community's data (see also section 5.3).

We note that this arrangement of Hosts tracking data producer and consumer applications and Edge Nodes tracking data producer and consumer Hosts in different Communities across separate Valleys and Namespaces is intended to provide a form of infrastructural aggregation that can benefit the Yodel architecture in several ways. First, it removes the need for Yodel Edge Nodes to contact the network controller for every Join the Community request they receive as long as they have prior records of data production or consumption for the given Community. Second, the network becomes more agile and faster to respond to changes in data production and consumption as a controller visit is only necessary once for data production and once for data consumption in a Community by an Edge Node. Third, it prevents changes in the forwarding paths when new producer or consumer Hosts in a Community join or leave the Yodel Edge Node which hugely improves routing scalability.

\color{DarkOrchid}\subsubsection{Signaling The Network Controller}\color{Black}

If a Yodel Edge Node does not find prior records of data production or consumption for a given Community in its PRT or CRT tables respectively (depending on the type of the request), the Edge Node contacts the Yodel controller (i.e., Yodel Master) to obtain the Community's corresponding Channel ID by adding its own YNI to the request. This step is similar to how SDN openFlow switches act when they see a flow for the first time \cite{sdn-nick-2009} which as we mentioned before ony happens for each Yodel Edge Node once for observing the first data production and once for observing the first data consumption for a given Community. In response, the Yodel Master subcontroller will lookup its knowledge base to find prior records of the given Community in the intended VNS. As stated in section 4.2.3, individual Valleys have separate controller resources. That being said, Yodel Master holds and manage the knowledge of individual Valleys in separate instances of Valley Information Base (VIB).

If the Yodel Master subcontroller finds prior records of the given Community in the intended VNS by looking at the corresponding Valley's VIB, it then adds the Edge Node YNI and its role (i.e., producer or consumer Edge Node depending on whether this was a request for data production or consumption) to the Flow object that is associated with the given Community (see section 5.3.1) and returns a corresponding Channel ID to the requesting Edge Node. The Edge Node also creates a record for the given Community and Namespace in the FIB of the intended Valley and records the given Channel ID for the indented Community and Namespace as well as the YNI of the requesting Host in the associated Valley's PRT or CRT table depending on whether this was a data production or consumption request, as shown in Figure 13. The Edge Node will also inform the Host of the ID of the Community's Channel which the Host will also record and share with the corresponding data producer/consumer application for future use, as shown in Figure 13. We note that if the Yodel Master subcontroller cannot find records of the given Community in the intended VNS, the controller must first create a record for the given Community in the Community's associated Valley Information Base. Following that, it creates a Flow object (see section 5.3.1) for the given Community and a Channel ID that belongs to the Community for the first time. Then, the Yodel Master adds the corresponding Edge Node and its role (i.e., producer or consumer Edge) to the list of Flow's Edge Nodes and respond to the given Edge Node by sending the Edge Node the Channel ID of the corresponding Community which the Edge Node also records and shares with the requesting Host, and subsequently with the data producer/consumer applications for future use, as shown in Figure 13. 

\color{DarkOrchid}\subsection{In-Host and In-Edge Data Delivery}\color{Black}

The Joining Community process, outlined in the preceding section, is an iterative process applied to each distinct Community in the network. The process continuously adds data producers and consumers to their Communities and ensures that Edge Nodes, serving as Yodel producer and consumer Edges, Hosts, and data producers/consumers in all Communities to have the Community's Channel ID. As we mentioned earlier in section 5.1.1, Channel IDs are very important and directly map to distinct Communities. Basically, a Community's Channel ID is unique within each Valley and can be used by data producers and consumers to interact in a Community, and by Hosts and Yodel Edge Nodes to recognize and guide distinct Community's traffic in separate Valleys. In this section, we show how the Channel ID can be used for data delivery by Hosts and Yodel Edge Nodes.

\begin{figure*}[h!]
  \centering
  \animategraphics[autoplay,loop,width=16cm,height=5.75cm]{2}{images/delivery/delivery-}{0}{18}\bf\color{DarkOrchid}
  \caption{\color{Black}\small End-to-End data delivery in a Yodel Community. For simplicity a single data producer is used. This diagram is animated. (YCD= Yodel Client Daemon, FIB= Forwarding Information Base, P=data producer, and C= data consumer).}
\end{figure*}

\color{DarkOrchid}\subsubsection{In-Host Data Delivery}\color{Black}

In-Host data delivery is an stage in the delivery of a Community's named data objects to the Community's data consumers that involves the Hosts to process the Yodel L3.5 data messages and make forwarding decisions. This stage happens when 1) data producers and consumers running on the same Host must exchange data messages (i.e., interprocess communication), and 2) when producers data is received by a consumer Host from the Yodel Edge Node where it is connected to, as shown in Figure 14. All data producers include an intended Valley name/ID and an intended Community's Channel ID in the header of every data message they send out. As stated in section 5.1.1, Hosts record the application ID of the data consumers along with the corresponding Valley and Channel IDs of the associated Communities where they are registered in the CRT table. 

By comparing the information stored in the CRT table and the information stored in the producers data messages header (i.e., Valley and Channel IDs), a Host is capable of locating the applications IDs of the Community's data consumers running on the same Host for delivering the producers data (see Figure 14). We note that there are other pieces of information that a Host stores in the CRT table like Timer and Lock (see Figure 14) which are configured based on the specifications of the given Community's multicast service model. Basically, the timer is used to manage active Community registration periods and a lock is used to prevent a data consumer to receive producers data messages. We also note that the same pieces of information (i.e., application ID, Valley ID, Channel ID, Timer, and Lock) are stored for data producers registering in different Communities in the PRT table, as shown in Figure 14. The PRT table and the information stored in the table is also used by the Yodel architecture to control data production at the source (if needed), and based on the specifications of Yodel multicast services used by distinct Communities. 

\color{DarkOrchid}\subsubsection{In-Edge Data Delivery}\color{Black}

In-Edge data delivery is an stage in the delivery of a Community's named data objects to the Community's data consumers that involves the Yodel Edge Nodes to process the Yodel L3.5 data messages and make forwarding decisions. This stage happens in two scenarios of: 1) moving the Community's data between data producer and consumer Hosts connected to the same Edge Node, and 2) when a consumer Edge Node in the Community receives producers data from other producer Edge Nodes in the Community, as shown in Figure 14. Similar to Hosts, Yodel Edge Nodes also record the Communities Channel IDs and associated consumer Hosts in the PCT table (i.e., in a separate FIB instance for each Valley). Once a data producer generates a data message, the associated Host where the data producer is linked sends a copy of that message to the Yodel Edge Node where the Host is connected in a Yodel Point-to-Point (YPP) message format. The YPP message header contains the Valley ID/name and the Channel ID of the intended Community that this data message is programmed for delivery. The Yodel Edge Node then uses the message's Valley ID/name to lookup the Channel ID of the corresponding Community in the PCT table of the given Valley's FIB instance. We note that similar to Yodel's way of using Locks in PRT and CRT tables for controlling data production and consumption in Hosts, Locks are also used by Yodel in Edge Nodes' PPT and PCT tables for managing the Edge Nodes activeness for data production or consumption in various Communities. 

Therefore, if an active record is found in the PCT table, the Edge Node forwards a copy of the data message it received from the producer Host to other consumer Hosts in the given Community that are listed in the given record using the YPP message format (see Figure 14). Although, the same steps are followed by a Yodel consumer Edge Node when it receives producer data messages from other producer Edge Node in the network, such delivery also requires additional steps to process a different data message format that is described in section 5.4.2. We note that an Edge Node never sends a copy of a data message back to a producer Host in case that the producer Host is also listed as a consumer Host on the records (i.e., the Host has both data producers and consumers in the Community) as the data can be delivered to the data consumers on the same producer Host via the in-Host data delivery stage. Similarly, once a data message that the Yodel Edge Node sent out is received by consumer Hosts, they follow the in-Host data delivery process for delivering them to the corresponding data consumers.

\newpage

\color{DarkOrchid}\subsubsection{The Yodel YPP Message Format}\color{Black}

As mentioned earlier, moving data between producer and consumer Hosts that are connected to the same Yodel Edge Node happens using the Yodel-Point-to-Point (YPP) message format. Yodel also uses YPP messages for carrying control messages between Hosts and the Yodel Edge Nodes (e.g., Join Community messages).

\begin{figure}[h!]
 \includegraphics[height=3.55cm,width=5.75cm]{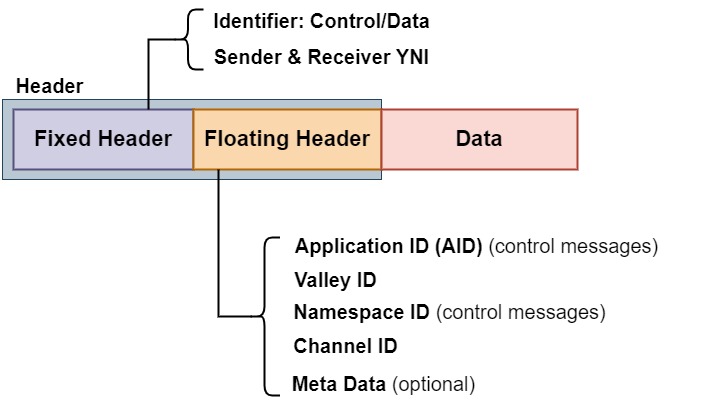}\bf\color{DarkOrchid}
  \centering
   \caption{\color{Black}\small A visualization of Yodel YPP Message Format.}
\end{figure}

The Yodel YPP message format, as shown in Figure 15, is composed of a header part and a data part. The header itself is composed of a fixed header and a floating header where the fixed header carries a message identifier indicating that the message is used for control or data delivery purposes, and also contains the YNI of the message's sender and receiver that are the endpoints of the communication. We note that since YPP messages are point-to-point, the receiver is always the message next hop and destination. The floating header also contains some important header elements, as shown in Figure 15, which can change depending on the type of the message (i.e., control vs data). For instance, the Applications and Namespace IDs are only used in the header of control messages for signaling the Edge Nodes during the Join Community process. For data production and forwarding purposes, all applications, Hosts, and Edge Nodes use the associated Community's Channel ID which is unique for the given Community regardless of which Namespace the Community belongs to in the intended Valley. This arrangement allows Yodel to reduce the number of header elements needed to be included in Yodel data messages which subsequently reduces the message processing time, bandwidth requirements, and latency of data delivery in Yodel Communities at scales without compromising the Yodel Conceptual Model requirements explained in section 4.2. We note that the number of data messages is way more than the number of control messages in the network. Meta Data in the floating header can also change depending on the type of the YPP message (e.g., attributes to describe the data object in a data message).

\color{DarkOrchid}\subsection{In-Network Data Delivery: Routing}\color{Black}

As we discussed in section 5.2, in-Host and in-Edge data deliveries are carried in a sequence of pre-determined point-to-point steps that do not require layer 3.5 routing. Basically, these stages are only forwarding steps guided by the use of Valley and Channel IDs. Nonetheless, there are situations where data consumers can Join a Community from Hosts that are connected to Yodel Edge Nodes other than the Community's producer Edges, as shown in Figure 14. In those situations, data messages carrying the Community's data must be forwarded through the Yodel Transit Network that interconnects Yodel Edge Nodes across multiple infrastructure domains. We refer to this stage as In-Network data delivery. As we know, forwarding Community's data through the Yodel Transit Network is not a one-hop point-to-point communication stage. Instead, data objects must pass through one or more Connector Nodes and infrastructure domains that require layer 3.5 routing for data delivery. Hence, In-Network data delivery involves Connector Nodes to process the Yodel L3.5 data messages and make forwarding decisions. We note that moving data between producer and consumer Edge Nodes in the same domain can be done directly without needing to pass through Connector Nodes. Nonetheless, the logic stays the same.

\color{DarkOrchid}\subsubsection{Routing And Forwarding Management Objects}\color{Black}

Routing of the Community's data is the job of the Yodel Master. As stated in section 4.2.1, distinct Communities in Yodel can use different multicast service models for communicating data over the same Yodel multi-domain network infrastructure. Hence, In-Network routing in between producer and consumer Edge Nodes must be performed separately for individual Yodel Communities and aware of the requirements of each Community's multicast service model and quality of service requirements. As discussed in section 3.3, rendering multiple multicast services over the same networking infrastructure (e.g., PIM-SSM \cite{ssm-bhat-2003} and PIM-SM \cite{pim_sm_motivation-deering-1996} in IP Multicast) can become very complicated for a network architecture to manage, mainly due to the differences in goals defined by distinct multicast service models. To reduce the complexity and remaining architecturally flexible to future changes (e.g., supporting additional multicast service models or changing the specifications of the existing ones), we designed Yodel routing mechanism with two major considerations in mind. 

First, the Yodel Master applies all constraints regarding the number of active data producers and consumers and subsequently the number of active producer and consumer Edge Nodes that can exist in a Community's Flow object at the time of Joining the Community, as defined by the Community's multicast service model, and before routing actually happens. There is only minimal changes that may happen afterwards which depends on the Community's multicast service model. Therefore, the routing function can assume that routing can be performed between any active producer and consumer Edge Node in the Community that is part of the Community's Flow object. We note that a big part of the differences between distinct multicast service models in Yodel lies in the number of active data producers and consumers in the Community.

Second, the Yodel Master uses a set of management objects, called Flow, Channel, and Path objects (see Figure 3) that together aim to help Yodel in two ways. First, they simplify routing and making re-routing much easier and less demanding by enabling Yodel to manage routing in multiple stages using divide and conquering methodology. In particular, Flow, Channel, and Path objects extend the divide and conquering approach used in Yodel conceptual model to the finest details of managing routing and forwarding in Yodel. Second, Flow, Channel, and Path objects provide a unified framework to perform source routing for all multicast services regardless of their differences. Such framework can hugely reduce the complexity of routing function in Yodel and, by extension, in name-based networks. In the following, we describe Flow, Channel, and Path objects, how they perform source routing, and the relations they have to each other to achieve divide and conquering.


\begin{itemize}[leftmargin=0.3cm]
    \item \textbf{Flow}: A Flow object, as shown in Figure 3, is a bipartite graph of producer and consumer Edge Nodes for a Community. As Edge Nodes can concurrently operate in both producer and consumer capacities, an Edge Node performing both roles is indicated in a Flow's graph as distinct nodes, one representing the producer role and the other one representing the consumer role (see Figure 3). A Flow graph also has a type that indicates the multicast service model that is used by the Flow's Community. Based on the Flow's type and when there is at least one active producer and one active consumer Edge Nodes in the Flow, provided that the two Edge Nodes are distinct, a Flow object may spawn one or more Channel objects. 
    
    \setlength{\parindent}{15pt} As stated in section 5.1, the Flow object is created and dynamically updated by Yodel Master during the Join Community process. Once a Flow object is created, a Channel ID will be also created with it as mentioned in section 5.1. This Channel ID is generated regardless of whether a Channel object will exist in the future, and will be assigned to the first Channel object once created. A separate Channel ID will be also generated for each additional Channel object that will be created and advertised to Yodel Edge Nodes, Hosts, and data producers/consumers in the Community. We also note that Channels and Channel IDs are different architectural resources. A Channel object must exist for In-Network data delivery to happen.

    \item \textbf{Channel}: A Channel object, as shown in Figure 3, is a graph consisting of a Flow’s Edge Nodes and associated set of Connector Nodes that interconnect the producer and consumer Edge Nodes across multiple domains. Depending on the Flow's type, a Channel can be single or multi source and may include all or part of the Flow's consumer Edges. The graph of Connector Nodes in the Channel is undirected as its links can carry traffic in one, the opposite, or both directions. Depending on the number of Channel's producer Edges, a Channel object may spawn into one or more directed Path objects that each carry the traffic (i.e., Community's named data objects) from a producer Edge Node in the Channel to every other consumer Edges. 

    \begin{figure}[h!]
        \includegraphics[height=3.5cm,width=7.75cm]{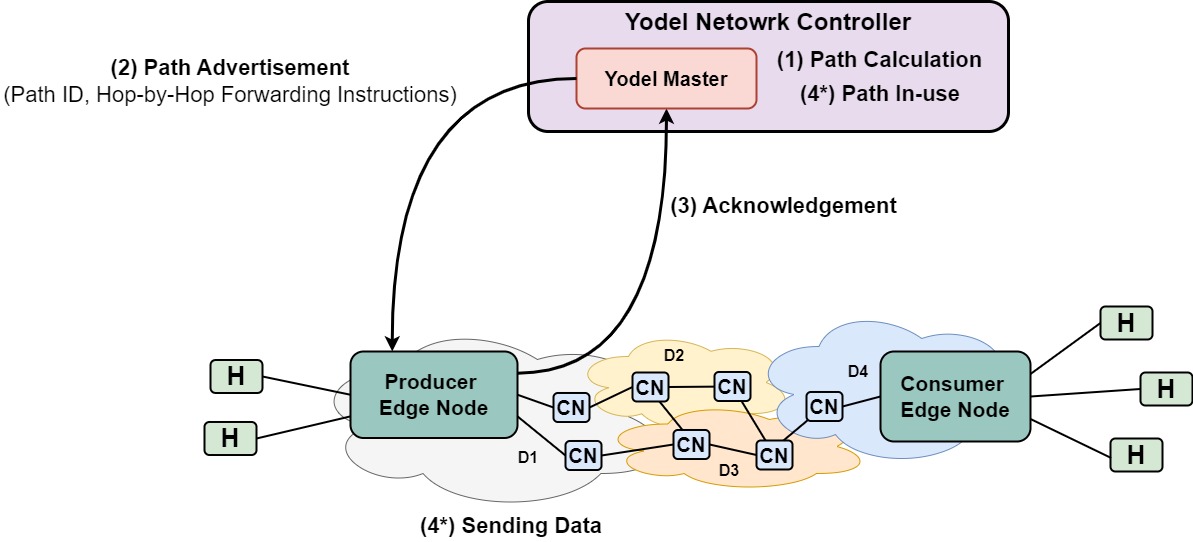}\bf\color{DarkOrchid}
        \centering
        \caption{\color{Black}\small Path Advertisement in Yodel.}
    \end{figure}

    \item \textbf{Path}: A Path object, as shown in Figure 3, is a unidirectional tree graph rooted at a Channel's producer Edge Node that sequentially traverses the Channel's Connector Nodes to move Community's data objects to the Channel's consumer Edges (i.e., in parallel). A Path is loop-free and is computed by the Yodel routing function from the information stored in the Community's Channel object. The Yodel routing function is part of the Yodel Master subcontroller. Once a Path object is created for a producer Edge Node in a Channel (i.e., the Path source), the Yodel Master informs the given producer Edge node of the computed forwarding path, as shown in Figure 16. At this point, the producer Edge Node has the entire forwarding path to send the Community's data producer messages it receives from its producer Hosts to other consumer Edge Nodes in the Community for final delivery to the Community's data consumers. We note that Paths are subject to change as Flow and Channel’s composition may change over time. For efficiency, Path objects can be computed proactively to reduce the delay of path calculations. 
\end{itemize}    

\color{DarkOrchid}\subsection{In-Network Data Delivery: Forwarding}\color{Black}

As stated in the previous section, an advertised Path object contains an entire forwarding path information from a producer Edge Node (i.e., the Path source) to all distinct consumer Edges in a Community with hop-by-hop instructions of how to move data messages through all intermediary Connector Nodes that exist between the given producer and consumer Edges across the entire Yodel multi-domain network infrastructure, as shown in Figure 14. Such ability gives Yodel a complete control over the forwarding paths that move data messages in separate Communities and it can reduce the delay of routing named data messages in the network. In the following, we first review key existing technologies that can be used for handling the forwarding of data in source routing networks. We then explain how the Yodel architecture processes forwarding path information in more details.

\begin{itemize}[leftmargin=0.3cm]
    \item \textbf{Bloom Filter}: is a hash-based probabilistic data structure that can present a set of objects in such a way that simple but fast membership queries can determine whether an object is a member of a given set \cite{bloom_intro-bluestein-2002}. Bloom Filters are used in Line Speed Publish Subscribe Internet Networking (LIPSIN) architecture \cite{lipsin-jokela-2009} to provide a L2/L2.5 name-based forwarding fabric. LIPSIN defines the entire forwarding path from a data producer to all data consumers in the network as a set of hop-by-hop network interfaces that are hashed by a Bloom Filter \cite{lipsin-jokela-2009}. The hash can then be included in the data packets header before the producer transmits them to the network. Every forwarder node in the network can then matches its interfaces against the hashed forwarding path information using a membership function (i.e., binary AND). Selected interfaces in each node can then move data packets forward. Several ICN architectures like PSIRP \cite{psirp-fotiou-2012} and PURSUIT \cite{pursuit-trossen-2012} also leverage LIPSIN to provide global ICN services. We note that Bloom Filters can provide a stateless method of forwarding named data packets in the network which can remove the need for holding network state information like data names on individual content routers. Unfortunately, Bloom Filters are subject to false positives which can be problematic when used for forwarding data packets in the network \cite{bloomfilter_false-tapolcai-2012}. In an ICN network, false positives may cause unwanted data delivery to data consumers in the network or causing false access to network caches. A very interesting solution for preventing false positives is proposed and used in PURSUIT architecture which is also a proprietary technology. 

    \item \textbf{BIER}: The Bit Indexed Explicit Replication (BIER) is an approach of using binary forwarding information to achieve stateless forwarding \cite{bier_rfc8279-dolg-2017}. BIER uses Bit Indexing mechanism to encode forwarding information. Basically, each router in a BIER domain has a unique BitPosition (BP) or index. The BIER domain’s ingress routers (i.e., path source) create BitStrings wherein each bit indicates a router that must receive the data packets. The position of the bits also indicate the order they must receive it \cite{bier_rfc8279-dolg-2017, bier_rfc9262-eckert-2022}. Each BIER router that receives a data packet sends it to an interface that leads to the next BIER router based on the next bit position in the BitString until the data packets cross an egress BIER router. This way, BIER can create a stateless L2/L2.5 forwarding domain for ICN and non-ICN architectures \cite{bier_rfc8296-dolg-2018} that is not prone to false positives. BIER is currently maintained by IETF. 

    \item \textbf{Segment Routing}: Segment routing is a source routing method that involves storing the entire forwarding path for every data packet sent to the network in the packet’s header \cite{segment_routing-fils-2018}. Every forwarder node (i.e., segment ingress router) in the network can then read the packet’s header, pops (removes) the next forwarder node address/ID from the header, and forward the packet to the next forwarder node \cite{segment_routing-fils-2018}. Segment routing is generally not prone to false positives and can be implemented in Strict or Loose fashions \cite{segment_routing-fils-2018}. In the Strict fashion, the next forwarder node address/ID refers to a node that is directly connected to the previous node in the forwarding path. In contrast, in the Loose fashion, the subsequent forwarder node address/ID does not necessarily point to a directly connected forwarder node. Instead, it can denote any forwarder node in the network reachable from the preceding forwarder node. Both Strict and Loose approaches can remove the need to hold network state information on the forwarder nodes in the network with the Strict approach to provide a fully stateless forwarding operation \cite{segment_routing-fils-2018}. A major issue, however, is that forwarding path can be lengthy and including all the forwarding path information in every packets header that passes the network may reduce the effective network bandwidth and increase packet processing times.

    \setlength{\parindent}{15pt} Today, segment routing is combined with a variety of networking protocols and technologies to serve different application use cases \cite{segment_survey-ventre-2020}. SRv6 is an example of using segment routing in IPv6 networks \cite{srv6_rfc8986-leddy-2021} where subsequent IPv6 router address/ID in a forwarding path sequence can be embedded in an IPv6 header. Segment routing has also been used in Software-Defined Networking where the SDN controller can conduct source routing based on the network topology and inform the packet senders of the forwarding path sequences \cite{sdn_segement-abdullah-2018}. Combining SDN with segment routing can provide an agile and flexible mechanism for managing traffic in dynamic networks like Yodel \cite{sdn_segement-abdullah-2018}. 

    \item \textbf{SR-MPLS}: is a source routing method where segment routing is combined with MPLS label switching. In this approach forwarding paths for different packet destinations are calculated at the first MPLS router, called the headend router \cite{sr_mpls-bash-2019}. Using the calculated forwarding path information and by applying the labelling mechanism, the MPLS network will then program itself for moving data packets in the network based on the labels that indicate the forwarding pathway. Path computation can either be performed on the headend router or on an external PCE device that is used for managing the traffic \cite{sr_mpls-bash-2019}. The SR-MPLS approach can benefit from the scalability and performance of label switching offered by MPLS networks while leveraging source routing capabilities. 
\end{itemize}

\color{DarkOrchid}\subsubsection{The Yodel Approach}\color{Black}

Yodel also leverages segment routing to process the path forwarding information when data messages crossing the Yodel Transit Network, as shown in Figure 14. For this purpose, Yodel uses a secondary data message format, called the YSync. The YSync message format, as shown in Figure 17, is especially designed for carrying forwarding path information.

\begin{figure}[h!]
 \includegraphics[height=3.55cm,width=5.75cm]{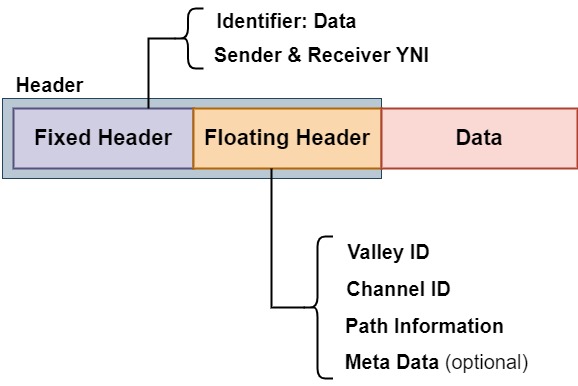}\bf\color{DarkOrchid}
  \centering
   \caption{\color{Black}\small A visualization of Yodel YSync Message Format.}
\end{figure}

The YSync message format is very similar to YPP with minor differences in the fixed and floating header elements. The fixed header, as shown in Figure 17, only indicates data type as YSync messages are not used for signaling and carrying control messages in the network. The floating header can also include the entire forwarding path information and data-message-only header elements, as shown in Figure 17. Once a producer Edge Node in a Community receives a Path object, it stores the Path information, along with the Channel ID that the Path advertisement belongs to in a table, called Active Forwarding Table or AFT, in the Edge Node's FIB, as shown in Figure 14. We note that similar to PPT and PCT tables, AFT tables are also stored in separate FIB instances that pertain to distinct Valleys. The information about the Valley ID for a Path object is also provided in the Path advertisement message.

The forwarding path information stored in the AFT table is used when a producer Edge Node in a Community receives data messages from the Community's producer Hosts that are connected to the Edge Node. Upon receiving the producers data messages, the Edge Node reads the header of the YPP messages that carry the producers data and use the Valley and Channel IDs stored in the messages floating header to lookup the AFT table and find the corresponding forwarding path information for each message. If a record is found, the Yodel Edge Node sends a copy of the data message in YSync format to other consumer Edge Nodes in the Community, as shown in Figure 14. We note that the consumer Hosts in the Community that are connected to the same Edge Node receive the producers data messages through In-Edge data delivery, as discussed in section 5.2.2.

The YSync data messages that are sent to the network carry the entire hop-by-hop forwarding path information for guiding the delivery of the messages to other consumer Edge Nodes in the Community. Therefore, each Yodel Connector Node in the network that receives a data message, reads the message header, pops (removes) the next Connector Node YNI(s) from the floating header, and moves the data message to the next Connector Node(s) in the downstream forwarding path. This cycle repeats by every Connector Node in the forwarding path until the message is finally delivered to all consumer Edges in the Community for final delivery to the Community's consumer Hosts and applications, as shown in Figure 14. We note that the message format changes from YSync to YPP for delivery to consumer Hosts after the message is received by a consumer Edge Node in the Community.

It is very important to mention that the organization of Yodel Proxy in Connector Nodes are different from Hosts and the Edge Nodes, and they process Yodel L3.5 messages differently. In particular, Yodel Connector Nodes do not record Valley, Namespace, and Channel IDs. They only store the information (e.g., YNI, link status information, etc) of their neighbors in a table, called Active Connection Table or ACT which we will describe in more details in the next section. The neighbor information enables Connector Nodes to move data messages forward in a forwarding pathway without needing to hold additional information. In other words, leveraging segment routing has removed the need for Yodel Connector Nodes to hold any Community-specific information as YSync messages carry the forwarding information (i.e., next Node YNIs) themselves. 

\color{DarkOrchid}\subsubsection{Yodel Strategic Forwarding}\color{Black}

\begin{figure*}[h]
  \includegraphics[height=6.75cm, width=15.5cm]{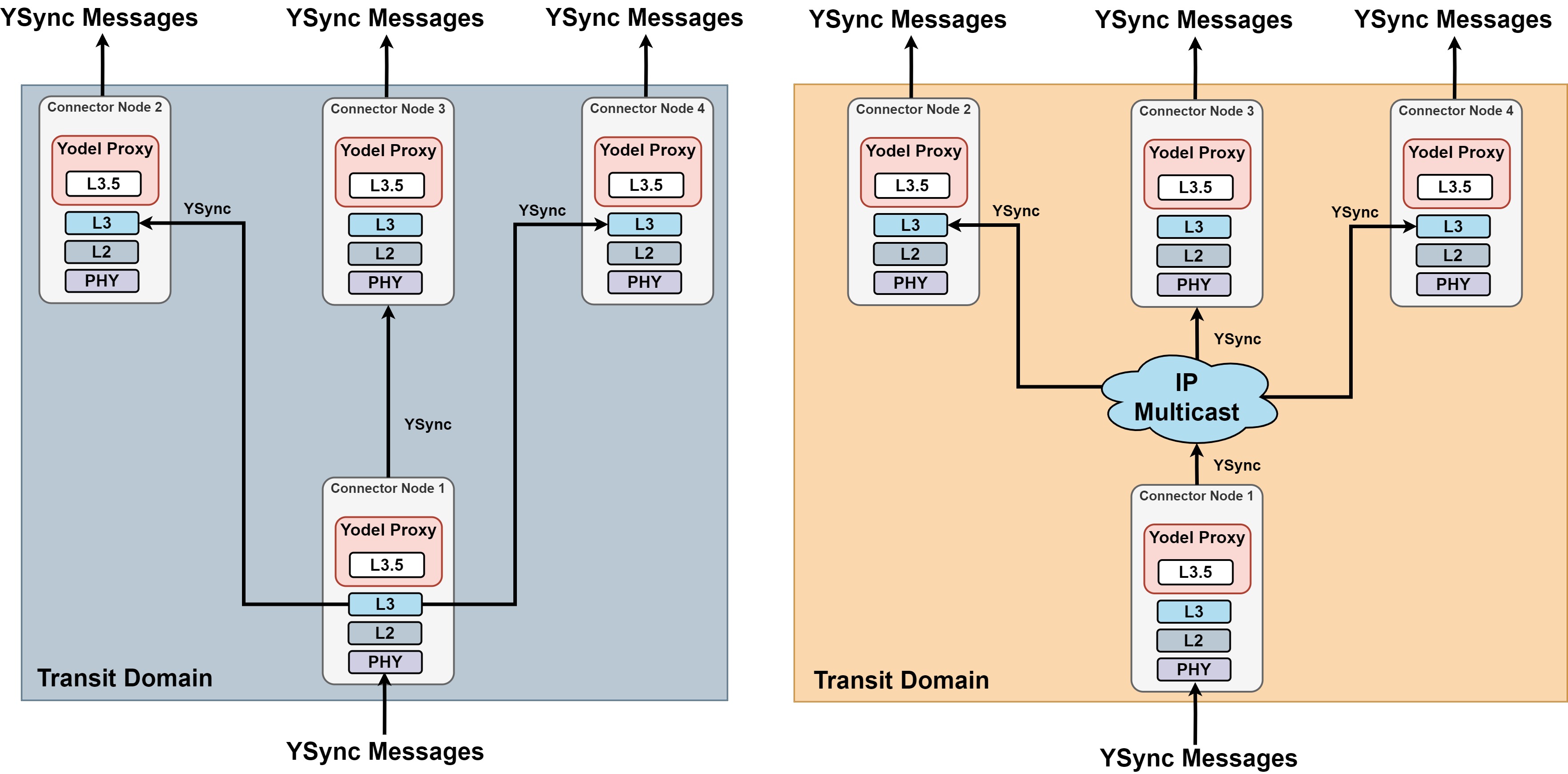}\bf\color{DarkOrchid}
  \centering
  \caption{\color{Black}\small The Yodel Strategic Forwarding is shown, choosing between Parallel Unicast (left) and IP Multicast (right) communication strategies to communicate data messages between the ingress Connector Node 1 and the three subsequent Yodel Connector Nodes in a transit domain.}
\end{figure*}

As stated in the previous section, Yodel Connector Nodes use ACT tables to move data to their neighbors, as shown in Figure 14. The table holds two pieces of information for each neighbor node: 1) the YNI of the neighbor node, and 2) the list of Active Forwarding Strategies (AFT) that is available to a Connector Node to select from and use to forward data messages to its neighbor node(s) in a forwarding pathway. The list of Active Forwarding Strategies (AFS) in the ACT table is prepared and dynamically changed for each neighbor node by the Yodel Proxy. Basically, each Forwarding Strategy is a list of downstream YNIs and their associated endpoint addresses from the layers below (e.g., IP address, MAC address, class D IP Multicast, etc.) as well as statistics related to the given strategy like its usage compared to other available strategies, latency, and availability. These statistics also change over time.

Let's see the example shown in Figure 18 for an explanation of how Strategic Forwarding works. In this example, the Yodel Connector Node 1 must transmit the incoming data messages to three downstream Connector Nodes. To do so, the Yodel Connector Node 1 has the ability to transmit data messages in two ways of: 1) using separate layer 3 unicast communications (Figure 18, left), or 2) reaching to the downstream Connector Nodes using an IP Multicast network that is reachable to all four nodes (Figure 18, right). Depending on the condition of the network within the given domain (e.g., link conditions, available bandwidth, quality of service requirements, etc.), the Yodel Proxy may choose either of the two strategies to move data to the downstream Connector Nodes. It is also possible for the Yodel Connector Node 1 to reach to two of the downstream nodes using the IP Multicast deployment, and one node, using a single one-to-one communication depending on the network and links conditions.

Once a data message arrives at the Connector Node 1, the node knows the subsequent (neighbor) nodes YNIs in the downstream forwarding pathway from the information stored in the message header. Based on the neighbor YNIs, the Connector Node can find the set of strategies to reach to each subsequent Connector Nodes in the ACT table and choose the (best) strategy that is available to move data to each node. In the case of choosing the IP Multicast strategy, a single strategy can move the data to all three subsequent Connector Nodes which uses the available network bandwidth more effectively as less messages are sent to the network. We note that the Yodel Proxy on Connector Node 1 can also be programmed to schedule or prioritize data delivery to individual downstream Connector Nodes, depending on the underlay network condition that enables the two node to communicate. It is also noteworthy to know that communications between any two nodes in Yodel L3.5 is connection-less.

A key benefit of performing Strategic Forwarding in Yodel is to improve the network resiliency and fault-tolerance. In other words, while segment routing removes Community-related forwarding information from the Yodel Connector Nodes for fast and stateless forwarding of Community's data messages, the Strategic Forwarding allow Connector Nodes to keep network-level state information that can enable the network to quickly recover from link or node failures. Strategic Forwarding also helps Yodel Connector Nodes to have the opportunity to use multipath forwarding for delivery of separate data messages in a Community to a downstream Connector Node using different communication links. 

The same as Yodel Connector Nodes, Hosts and Yodel Edge Nodes also use the ACT table and Strategic Forwarding method for transmitting Community's data with some differences introduced in the process. For Yodel Edge Nodes, processing ACT table and strategy selection happens after looking up the PCT and AFT tables for finding subsequent consumer Hosts and Edge Nodes for delivery of a data message in the network. Once all recipients of a data message is known to the Edge Node, the Edge Node proceed to forwarding with looking up the ACT table to find the (best) strategy to move data to subsequent nodes. A Yodel Edge Node also uses its ACT table when responding to the Hosts control messages. For Hosts, processing ACT table happens under two circumstances. First, when data messages in Community are delivered to data consumers running on the Host. In this situation, the Host finds the recipient data consumers by first processing the CRT table. Once the receiver data consumers are known to the Host, the Yodel Proxy on the Hosts lookup the (best) strategies available on the ACT table to forward data messages to the given data consumers. We note that data producers/consumers can link to a Host through various methods available through the Host's YCD like Unix Sockets, Named Pipes, and UDP. Using Named Pipes and shared memory models, Yodel Hosts can achieve multicast data delivery to a group of intended data consumers on the same Host. Host's Yodel Proxy also looks up ACT table for finding (best) forwarding strategies for every data and control message that it communicated with an Edge Node.

%% file: content/dt.tex
\color{RedViolet}\section{Connection Resiliency}\color{Black}

\begin{figure*}[h]
  \includegraphics[height=6.25cm, width=14.5cm]{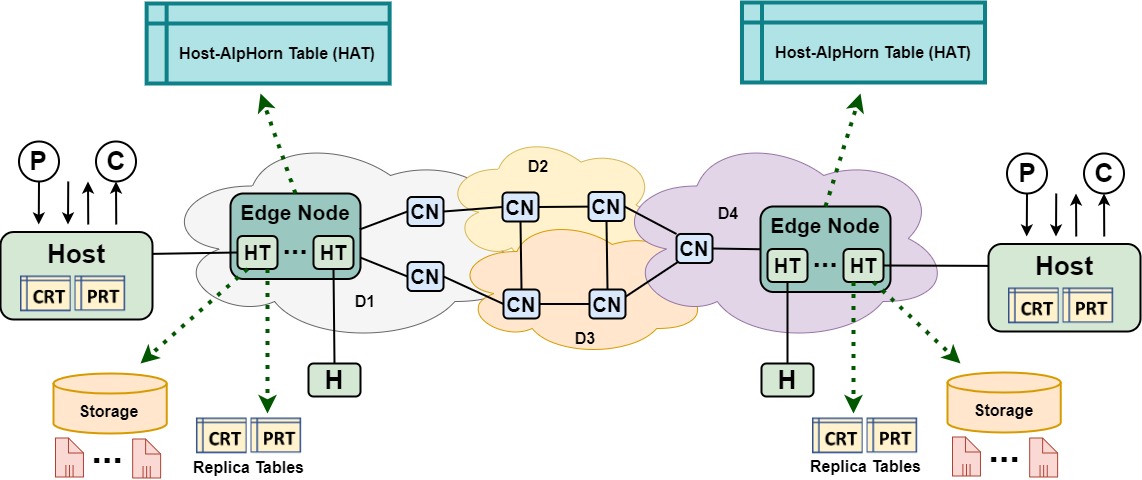}\bf\color{DarkOrchid}
  \centering
  \caption{\color{Black}\small The Yodel Host Twin object is shown with several in-network components that Yodel uses in Edge Nodes to support connection resiliency. (H=Host, HT=Host Twin, P=Data Producer, C=Data Consumer, and CN=Connector Node)}
\end{figure*}

As we explained earlier in section 3.5, a key Yodel design goal is to improve its data delivery capabilities with additional support for applications under temporary Host disconnection. Basically, Yodel uses an in-network digital twin object, called the Host Twin \cite{yodel-morteza-2023}. A Host Twin object can become active once its corresponding Host disconnects from the network (subject to timeouts). During this time, the Host Twin object performs two essential tasks. First, it keeps all data producers and consumers registration in different Communities active so that when the Host connectivity resumes (subject to timeouts), the Host's applications can seamlessly continue to send and receive data objects within the Communities they are registered in. Second, the Host Twin object continues to receive data objects on behalf of its corresponding Host's data consumers until the given Host connectivity resumes or the Host Twin timer expires. Once the given Host reconnects with the network, the Host Twin object becomes inactive and the Host's data consumers can access the collected data during the disconnectivity period. We note that as discussed earlier, data producers and consumers on a Host are linked to the Host's YCD and interact with the network and Edge Nodes through the corresponding Host. Therefore, there is no session between the data producers/consumers on the Host and the Edge Nodes in the network that requires maintaining session state information to resume connectivity. If a Host reconnects to the network, data producers and consumers can continue working as long as their Community registrations remain active for which we use the Host Twin object. We also note that the fundamental assumption here is that the underlying causes of Hosts disconnections are link failures and network unavailability not Hosts hardware malfunctioning which requires rebooting the Host and running user programs again. 

\color{DarkOrchid}\subsection{The Host Twin Object (The AlpHorn)}\color{Black}

The Host Twin (HT) object, also called AlpHorn, is a self-contained virtual compute environment that is created and managed by a Yodel Edge Node, as shown in Figure 19. Yodel Edge Nodes create Host Twin objects during the Host provisioning process, explained in section 4.3.3. Basically, once a Host is connected with a Yodel Edge Node, the Edge Node creates a Host Twin for the connected Host and configures the Host Twin object to interact with the Host in order to collect important information like Host responsiveness and reachability. The Host Twin also collect information regarding the Host's data producers and consumers registration in different Valleys and Communities. To do so, the Host and its associated Host Twin object regularly interact to keep an up-to-date replica of the Host's CRT and PRT tables on the Host Twin object, as shown in Figure 19. Communication between a Host and its Host Twin object (AlpHorn) is query based where the query and its response are encapsulated in YPP control messages without using a floating header. It is just the fixed header and the data part where the query and its response are coded.

\color{DarkOrchid}\subsection{Managing Host Twin Objects}\color{Black}

As we mentioned in the previous section, management of Host Twin objects is the responsibility of Yodel Edge Nodes. The entire process of creating, managing, and destroying Host Twin objects are handled locally on each Edge Node and does not involve the network controller. This is particularly a very important design decision that helps Yodel to realize its connection resiliency goal while not sacrificing performance achievements by involving visits to the network controller. We note that actual process of Host Twins instantiation is a very detailed process that involves the use of container management and orchestration systems like Kubernetes which we believe deserves an independent discussion in a separate paper.

To maintain records of Host Twins on the Edge Node, each Yodel Edge Node records all connected Hosts and their associated Host Twins in a table called the Host-AlpHorn Table or HAT. At a basic level, HAT stores the Host YNI, the AlpHorn (Host Twin) ID, and a timer for the Host Twin object. The AlpHorn ID is a locally unique YNI that the Edge Node creates and associate with the AlpHorn/Host Twin object after it is instantiated. This ID will be also shared with the corresponding Host and is stored on the Host's ACT table. This way, Hosts always recognize their Host Twin object as a neighbor Node. The Yodel Edge Nodes also store AlpHorn IDs (YNIs) in their ACT table, so that they can pick (the best) strategy for delivering the message to the intended Host Twin object once a control message is received. We note that using a YNI (i.e., layer 3.5 address) to identify Host Twin objects allows Yodel to handle Host-to-AlpHorn (Host Twin) communications in layer 3.5. The timer on the HAT table is also used to avoid zombie AlpHorns to remain active in situations where a Host disconnects and does not reconnect afterwards. In this situations, the Yodel Edge Node destroys the corresponding Host Twin object of the disconnected Host when the timer expires. The timer value is settable.

\color{DarkOrchid}\subsection{Delivering Data To Host Twin Objects}\color{Black}

As we mentioned in the previous section, a Host Twin object of a connected Host collects status information about its corresponding Host to understand whether the Host is functional and reachable. In case, a Host is not reachable (over a settable period of time), the Host Twin object of the Host becomes active. In this situation, the Yodel Edge Node replaces the Host YNI with the Host Twin YNI in PCT table of every Valley's FIB instance. Yodel Edge Nodes typically index the Valleys where a Host Twin object has dependency and use the indexes to quickly replace the given YNIs. Once the YNIs are replaced, all data objects that are intended for delivery to the given Host are now being delivered to its Host Twin object instead. The Host Twin object also stores the objects in an Edge Node storage for future use. Once a Host reconnects, the Yodel Edge Node finds its YNI in the HAT table first to ensure that this is a returning Host. If the Host YNI is found, the Yodel Edge Node replaces the Host Twin object's YNI with the Host' YNI in the PCT table of every Valley's FIB instance the same way as we described the reverse. Afterwards, the Host can begin receiving data objects that are destined to the Host. The Host will also receive data objects that are collected by its Host Twin on the Edge Node. 

\color{DarkOrchid}\subsection{Benefits Of Digital Twinning For Yodel Multicast Services}\color{Black}

As we discussed, a key benefit of digital twinning in Yodel is to maintain in-network status about Hosts that are connected to the network and their applications so that the network can provide enhanced communication services like connection resiliency. Connection resiliency can play a significant role in using Yodel multicast services in environments where connectivity is not persist or highly available such as Internet of Things application environments. In many use cases, IoT applications demand connectivity with sensors and IoT devices that are in remote or rural places. Network connectivity in such environments is not always a commodity to rely on and therefore, Yodel's approach to digital twinning can improve applications performance and reliability. Digital twinning can also help many other Internet multicast applications like video conferencing with ways that allow them to benefit from historical data that have been collected during the time of disconnectivity. For instance, if a user's device in a video conferencing/call gets disconnected, the conferencing application on the user's device (e.g., laptop or mobile phone) can retrieve the data from the network once the user device reconnects, allowing the user to begin communicating with others in the conference/call while providing the user with a text version of the communications the user lost (i.e., prepared by the application) on a side bar note in real-time.

The Yodel's approach to leverage digital twinning can also help many AI and machine learning applications to avoid losing data during any unexpected disconnection times. For instance, federated learning applications may have heterogeneous user devices that are placed in different geographical locations with non-uniform access to Internet connectivity. Since collecting local data is very important for improving the local and global model's performance and accuracy, accessing the historical data collected during disconnection can help distributed agents to obtain the data after they recover from network interruptions and use the historical data to improve their models. Last but not the least in our mind is the ability to use Host Twins for empowering Yodel multicast services with in-edge processing of applications data. In this scenario, instead of always using Host Twin objects' computing power for data collection during Hosts disconnectivity, we can use Host Twins to pre-process data objects before delivery to their intended destinations. An example would be the data producers to generate text messages in a Community where each intended recipient can obtain the data translated in a language of interest. The translation can happen using Host Twins on the Edge Node where the intended consumer Hosts are connected before delivering data to the given Hosts. Pre-processing data objects can also help AI applications to clean data while it is in transit using any of th Yodel multicast services. Such pre-processing can happen using Host Twins capabilities on either or both of the producer and consumer Edge Nodes in a forwarding pathway.

%% file: content/summary.tex
\color{RedViolet} \section{Concluding Remarks And Future Work}\color{Black}

As we discussed throughout this paper, the current Internet architecture was initially designed to facilitate communication between any two endpoints in the network. This one-to-one form of communication has been very useful to enable Internet unicast applications like email, banking, and file transfer. With the advent of Web and Mobile platforms, new Internet applications have emerged that are inherently multicast-oriented, requiring endpoints to communicate in groups that necessitate one-to-many and many-to-many communication capabilities. Various proposals have emerged for adding multicasting capabilities to current Internet architecture like IP Multicast and MBone. However, several decades of researching into key challenges of integrating multicasting with current Internet architecture have not provided enough incentives for the Internet Service Providers (ISPs) to offer widespread multicast support, especially for the home users that encompass a large portion of the Internet users and multicast applications audience. The lack of technology support at the network-level has gradually pushed applications towards complex application-level solutions like Content Delivery and Peer-to-Peer network to meet their multicast needs. 

In the quest for innovations and alternative technologies to enable network-level multicast support, name-based networking (emerged through the advent of Information-Centric Networking) stands out. Name-based networking promotes an alternative and promising networking paradigm wherein communication is driven by names rather than addresses. In other words, the main emphasis is on \say{WHAT} data is being communicated in the network rather than \say{WHO} is communicating the data. Such small but fundamental shift in the networking paradigm has enabled name-based networking to improve network security by securing data objects instead of end-to-end channels, to enhance endpoint mobility (i.e., no change of addresses), and to natively support multicasting as multiple senders and receivers can communicate through a single name. To move towards a system of supporting name-based networking that is more aligned with multicast applications needs, we propose the Yodel architecture.

The Yodel architecture, as discussed in detail in this paper, is built upon several key design goals that allow the architecture to serve and meet the needs of multicast applications at multi-domain scales. In particular, Yodel emphasizes the use of name-group binding principle and push-based communication model as ways to improve routing scalability and simplifying communication within multicast groups. Multi-tenancy, as proposed by Yodel through its multi-layer conceptual model, can also improve privacy of communication, offer name reusability, and support multicast applications with a customizable networking experience. The Yodel multi-form multicast communication capability is also an step forward to empower Internet applications to further meet their multicast communication needs. Another step forward in this direction is the Yodel support of connection resiliency to help improving multicast applications experience by keeping them connected with network services and data in application environments where network connectivity is not always available. Finally, and most importantly, is the Yodel's approach to leverage layer 3.5 networking as a way to achieve three important goals: 1) removing the need for clean-slate infrastructural upgrades to deploy name-based routing and forwarding which has always been a preventive factor in large-scale deployment of ICN architectures, 2) providing backwards-compatibility by following the current Internet deployment pathway, and 3) to offer multi-domain networking without needing to change the existing internetworking infrastructures.

Moving forward, we are looking at several key aspects of improving Yodel implementation and enhancing its architecture. We have begun investigating Yodel implementation, how to build, test, debug, and deploy the architecture on Smart Application on Virtual Infrastructure (SAVI) testbed as its home \cite{savi-alg-2014}. So far, several key functions of the architecture have been prototyped and used in several IoT-related projects in Network Architecture Lab at University of Toronto \cite{yodel-morteza-2023}. In the future, we will extend collaborations with other researchers to explore various use cases of using Yodel in different application environments and extending our implementation efforts. In pursuit for architectural improvements, we are actively exploring ways of using digital twins for integrating Yodel's multicasting features with Edge and Cloud computing to support distributed content-oriented applications. Data and endpoints mobility and how we can leverage digital twinning to improve the network support of mobile applications is another aspect that we are currently investigating. Finally, we are also exploring how the Yodel digital twinning mechanism can help to add accountability to Yodel multicast services to provide economic benefits and incentives for Internet applications that require that to meet their business end goals.

%% file: content/media.tex
\color{RedViolet} \section{Guide For Reading This Paper}\color{Black}

This document contains many graphics which some are also animated. To clearly see the animations, you must download the PDF version of this document and open the PDF file in a compatible PDF reader. For your convenience, we have tested the animated PDF using Adobe Acrobat Reader version 2023.008.20470 and Foxit PDF Reader version 12. Please note that opening this PDF file in a browser does not execute animations probably. 

%% file: content/naming.tex
\color{RedViolet} \section{Why The Name Yodel?}\color{Black}

The history of telecommunication, the act of transmitting signals over a distance for communication purposes, spans thousands of years. Among the methods of communication, yodeling stands out. Yodeling is a distinctive form of singing characterized by rapid switches between low (chest register) and high (head register) pitches. This unique vocal style evolved into a longstanding rural tradition. Notably, yodeling has served practical communication purposes in Alpine regions, where mountaineers used it to communicate signals across vast valleys and distant villages. Herders in the same region also adopted yodeling as a means to call and manage their flocks. 

We initiated the design of the Yodel architecture within the IoT landscape where massive number of devices sense the environment for extracting actionable information. Within this context, multicasting sensor data has been always a key challenge, and we intended that our system would allow IoT devices to communicate as easily as a mountaineer can yodel into a Valley. As our concepts evolved, the Yodel architecture expanded its scope beyond IoT, catering to the multicast application needs in various domains. To pay homage to the tradition of yodeling as an analog method of telecommunication and to keep its spirit alive, we named our architecture Yodel. Consequently, many elements within the Yodel architecture, such as Valleys and AlpHorns, are named in harmony with this tradition.

%% file: content/service.tex
\color{RedViolet} \section{Yodel Multicast Services}\color{Black}

As explained in section 3.3, Yodel can provide multiple name-based multicast service models. These services control the number and the way that data producers and consumers can interact with each other within Yodel Communities. We believe that choosing a right service model for a Community requires a good understanding of how the network renders different multicast services in run-time. In section 4.4, we briefly introduced the Yodel multicast services. In section 5, we described how routing and forwarding happens in Yodel for a general multicast service model where multiple data producers can send their data to multiple data consumers in a Community. In this appendix section we study Yodel multicast services in more details. In particular, we: 1) describe individual service specifications and explain how they are different from the general service behavior we used in section 5 to explain routing and forwarding in Yodel, 2) explain the way individual services' Flow objects are managed, and 3) provide relevant use cases to better understand the nature of individual services. Table 1 below also provides a summary and comparison of all Yodel multicast services and their key attributes.

\begin{table}[h]
\bf\color{DarkOrchid}
\centering
\caption{\bf\color{Black} Comparison of Yodel Multicast Services}
\label{tab:services}
\scalebox{0.8}{
\normalfont\color{Black}
\begin{tblr}{
  cells = {c},
  cell{1}{1} = {r=2}{},
  cell{1}{2} = {c=4}{},
  vline{2} = {1}{},
  vline{2} = {2}{},
  vline{3-5} = {2}{},
  vline{2-5} = {3-9}{},
  hline{1,3-9} = {-}{},
  hline{2} = {2-5}{},
}
{\textbf{Yodel}\\\textbf{~Multicast Services}} & \textbf{ Attributes}                  &                         &               &              \\
& {\bf \# of Active\\~Producer Edges Nodes} & \bf \# of Channels Per Flow & \bf Channel Type  & \bf Partitioning \\
Yodel-SSM                                      & One                                   & One                     & Single Source & No           \\
Yodel-AC                                       & One                                   & One                     & Single Source & No           \\
Yodel-SLSM                                     & One or More                           & One or More             & Single Source & Yes          \\
Yodel-SLAC                                     & One or More                           & One or More             & Single Source & Yes          \\
Yodel-MSM                                      & One or More                           & One or More             & Multi Source  & No           \\
Yodel-MSAC                                     & One or More                           & One or More             & Multi Source  & No           \\
Yodel-MMM                                      & One or More                           & One or More             & Multi Source  & No           
\end{tblr}
}
\end{table}

\color{DarkOrchid} \subsection{The Yodel Single Source Multicast (Yodel-SSM) Service}\color{Black}

\color{DarkOrchid} \subsubsection{Key Service Specifications}\color{Black}

The Yodel-SSM service specifies a multicast communication model where only a single data producer in a Community can send its data messages to the entire Community's data consumers. The service allows more than one data producer to join a Community but all data producers must remain on-hold until the current Community's active data producer becomes inactive or unavailable (e.g., registration timer expires or the data producer disconnects). Keeping a Community's data producers on-hold happens by setting the corresponding Host's PRT table's Lock to one for the given data producers during \say{Join the Community} process. If a data producer Edges Node in a Community is locked during the Joining process, the Edge Node and all its data producers are locked until the lock is cleared. A producer Edge Node Lock in a Community can be set by adjusting the Edge's PPT table's Lock to one for the given Community. We note that the network controller (i.e., Yodel Master) can Lock any producer Edge Node in a Community's Flow object at any give time which subsequently Locks all data producers in the Community that connect to the network through the given producer Edge Node.

Unlocking data producers in a Community happens under the discretion of the Yodel Edge Nodes and the network controller (i.e., Yodel Master). If the current active data producer in the Community becomes inactive and there are existing on-hold data producers in the Community that are connected to the same Edge Node, the Edge Node can make a decision about which data producer to become active next. To do so, the Yodel Edge Node signals a (selected) producer Host in the Community to clear the Lock for a single data producer on the Host that belongs to the Community. If there is no other data producer in the Community that connects to the network through the same Yodel Edge Node, the Edge Node asks the network controller to remove the Edge Node as a producer Edge from the Community's Flow object where the network controller can subsequently unlock another producer Edge Node in the Community's Flow object (if any) to become active. Once the given producer Edge Node becomes active, it also clears the lock for a single data producer in its locality as described. We note that once a producer Edge Node removes itself (i.e., its role) from a Community's Flow object, all associated Path objects initiating from that producer Edge Node will be removed as well.

\color{DarkOrchid} \subsubsection{Service Flow Management}\color{Black}

\begin{figure}[h!]
    \includegraphics[height=4.5cm,width=4.25cm]{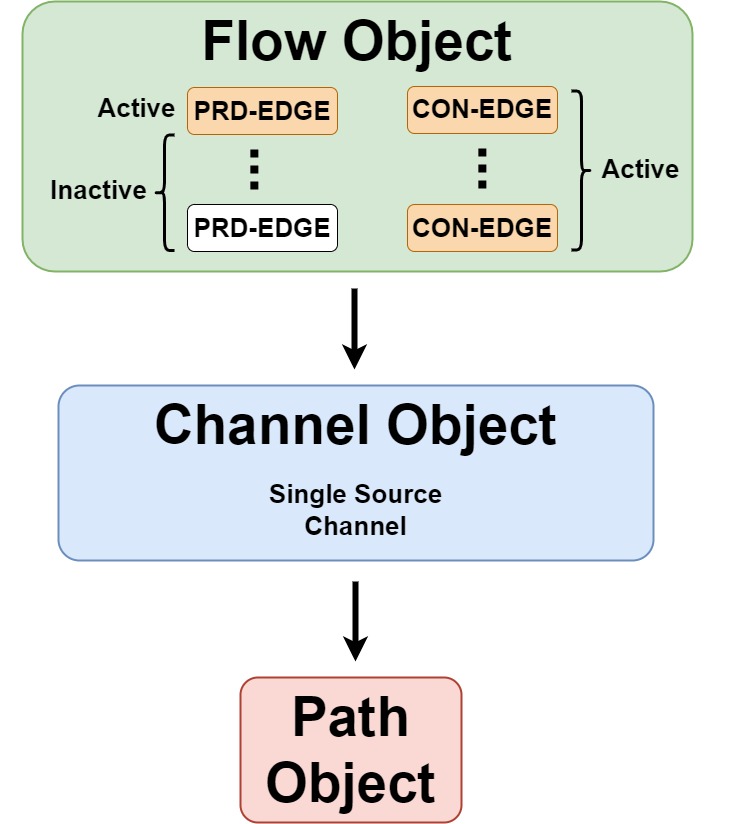}\bf\color{DarkOrchid}
    \centering
    \caption{\color{Black}\small Flow, Channel, and Path objects in Yodel-SSM service (PRD=Producer and CON=Consumer).}
\end{figure}

According to its service model, a Yodel-SSM Flow object can contain multiple data producer and consumer Edge Nodes for a Community where only one of the existing producer Edges in the Community is active at any time, as shown in Figure 20. Since the service model specifies the use of a single data producer in the Community, the service always has a single source Channel object, a single Channel ID that is assigned to the only service Channel object, and a single Path object that is created to move data between the only (active) producer Edge Node in the Channel to all other consumer Edges in the Community. We note that since the service Flow object has the knowledge of all inactive producer Edge Nodes in the Community, the network controller can pre-compute all the possible forwarding path configurations that begin with different producer Edges in the Flow. This way, the service response time improves drastically as changes happens in the Flow's active producer Edge Node.

\color{DarkOrchid} \subsubsection{Relevant Use Cases}\color{Black}

The Yodel-SSM service can benefit multicast applications in a variety of scenarios that involve distributing video, audio, and documents. For instance, Yodel-SSM service can be used to enable a network-level multicast experience for distributing online video streams like news broadcasts to its viewers. Distributing online video streams is typically done by relying on Web protocols like HTTPs and QUIC to carry the stream data while also leveraging Content Delivery Networks (CDNs) for content caching and application-level multicast support. Of course, IP Multicast can also provide network-level support in these scenarios, but the ongoing lack of global service availability makes it complicated to offer end-to-end multicast services, especially that multicast applications need multi-domain support. Using Yodel-SSM, multicasting data can instead become less complicated, seamless in multi-domain settings using Yodel layer 3.5 capabilities, and easily manageable.

\color{DarkOrchid} \subsection{The Yodel AnyCast (Yodel-AC) Service}\color{Black}
\color{DarkOrchid} \subsubsection{Key Service Specifications}\color{Black}

The Yodel-AC service builds on top of the Yodel-SSM service to enable multicast applications with enhanced control over the data consumers. Basically, the Yodel-AC service can let a subset of all data consumers in a Community to receive the Community's data messages which can happen in two ways of randomized or dedicated selection. In randomized selection, each of the producer Host, producer Edge, and Connector Nodes in the Community can choose to randomly move the Community's data messages forward. For instance, Hosts can only deliver a Community's data message to one of several data consumers in the Community that are running on the same Host. Similarly, Yodel Edge Nodes in the Community can choose to forward the Community's data messages to none, one, or multiple consumer Hosts that are connected to the same Edge Node. They can also choose to drop or move the Community's data messages to other consumer Edge Nodes in the Community but they can't modify the forwarding paths that are advertised to them for the given Community. Instead, Yodel Connector Nodes in the forwarding path will drop or move the Community's data messages forward to one or more nodes in the downstream forwarding pathway after popping the next node YNIs from the message header. Randomized forwarding of data messages in Yodel-AC Communities is achieved by data messages (i.e., both YPP and YSync) to carry a different data message identifier for the service than the one that other Yodel multicast services use.

To achieve a dedicated selection of recipients, individual data consumers in the Community must ask their Hosts to lock them in the Host's CRT table which prevents the Community's data messages to be delivered to them in the future. If all data consumers on Host a request the lock, the consumer Host also informs the Edge Node where it is connected to set a temporary data consumption lock for the Host in the given Community in the Edge's PCT table of the corresponding Valley where the Community belongs. We note that data consumers can also choose to remove the lock when they require to obtain the Community's data messages which subsequently can trigger the Host to remove the Host's lock on the Edge Node if one is already set for the Host. We also note that data consumers can only lock and unlock themselves when the Yodel-AC service is used in the Community. In all other services, locks are always set by the Yodel Edge Nodes and/or the network controller. It is also noteworthy to mention that there is no differences in how the Flow object is managed in Yodel-AC service compared to the Yodel-SSM service as the two share the single source multicast service model.

\color{DarkOrchid} \subsubsection{Relevant Use Cases}\color{Black}

The Yodel-AC service can benefit multicast applications in a variety of scenarios. For instance, the service dedicated mode can let data consumers to control when to receive data from the Community's data producer without needing to resubmit data consumption requests. Imagine, a data consumer that requires a document update every two hours while the updates are distributed to all data consumers from the source (i.e., data producer) every 5 minutes. So instead of needing the data consumer to resubmit a data consumption request each time, the data consumer can easily unlock itself every two hours which allows the data consumer to instantly receive an update once the data producer sends one to the Community. The data consumer can lock itself again for two hours following the receipt of the message. We note that locking and unlocking happens automatically and does not change the status of the network. Therefore, all forwarding paths remain established which can significantly reduce the overhead of reconfiguring the network each time a new data consumption request arrives. The randomization mode in Yodel-AC can also help multicast applications in event-driven systems to perform load-balancing as not all data instances need to be processed by all data consumers.

\color{DarkOrchid} \subsection{The Yodel Selective Source Multicast (Yodel-SLSM) Service}\color{Black}
\color{DarkOrchid} \subsubsection{Key Service Specifications}\color{Black}

The Yodel-SLSM provides a special multicast service model where instead of always keeping one data producer active in a Community as offered by the Yodel-SSM and Yodel-AC services, it partitions the Community into disjoint groups wherein each group can have a single active data producer sending its named data messages to the Community's data consumers. Partitions are driven either by the Yodel-SSM service, providing the Yodel-SLSM service, or by the Yodel-AC service, offering the Yodel-SLAC multicast service model. We note that the fundamental assumption in both the Yodel-SLSM and Yodel-SLAC services as well as the Yodel-SSM and Yodel-AC services is that all data producers (i.e., active or on-hold) in the Community provide exactly the same data objects to the Community's data consumers.

Partitioning in Yodel-SLSM/Yodel-SLAC services is the key service attribute. Basically partitioning in its current version begins when there is more than one data producer Edge Node in the Community's Flow object. Initially, subsequent data producer Edge Nodes joining the Community's Flow object are put on held (i.e., only the first producer Edge Node remains active). The network controller (i.e., Yodel Master) may begin to partition the Community after a second data producer Edge Node join the Community's Flow object. If the network controller partitions the Community, each partition gets a new Channel object and ID which the network controller advertises to all producer and consumer Edge Nodes in the Community that are now part of a new Channel object. The data produce Edge Node in the new partition will also receive a controller message instructing the Edge Node to become active. All Edge Node recipients of the new Channel ID in the new partition will replace the Channel ID they obtained before with their current Channel ID for the Community. The Yodel Edge Nodes also inform all Yodel Hosts in the Community of the new Channel ID which they subsequently inform their data producers/consumers in the Community of the new ID of the Community's Channel. This way all recipients of the new Channel ID become an entirely separate partition within the same Community. We note that one partition among all partitions can remain on the first Channel object and ID with no changes. Hence, producer/consumer Edge Nodes in that partition will never receive a new Channel ID advertisement. 

Once new Channel objects are created for new partitions, routing happens independently for each Channel object which also leads to the instantiation of separate Path objects within each partition. If changes happens in the Community's network such as a data producer Edge Node in a partition is removed from the Community's Flow object or the partition data consumers are not available anymore, the network controller may choose to unpartition an existing partition by merging it to another partition in the Community. If a merger happens, the producer and consumer Edge Nodes in the \say{set-to-be-merged} partition receive a new Channel ID which they update with their existing Channel ID for the Community. As mentioned earlier, this Channel ID is also shared with data producer/consumer Hosts and applications in the \say{set-to-be-merged} partition to update them with the new partition Channel ID. If there is only a single partition left in the Community, the partition can exist as long as the Community's Flow object exists.

We note that from time to time, the network controller may also reassign data producer and consumer Edge Nodes in a Community to new or existing partitions to optimize resource utilization, improve latency of data delivery, and enhance reliability of communication in the Community. We also note that separate partitions can sometimes include only a subset of all data producer and/or consumer Hosts of a Community that are connected to the same Edge Node. If that happens, the network controller provide the partitions Channel IDs to the given Edge Node where the data producer and consumer Hosts are connected and instruct the Edge Node to add its data producer and/or consumer Hosts to the new partitions at its discretion. In the future, we also aim to introduce high reliability partitions by allocating independent data producers to separate partitions. By independent data producer we mean that they fail in statistically independent manner. For instance, two data producers in the same Community that are connected through different Producer Edge Nodes can fail independently but if they connect to the network through the same Edge Node, the given Yodel Edge Node can become a common point of failure for both data producers.

\color{DarkOrchid} \subsubsection{Service Flow Management}\color{Black}

\begin{figure}[h!]
    \includegraphics[height=5.75cm,width=8.75cm]{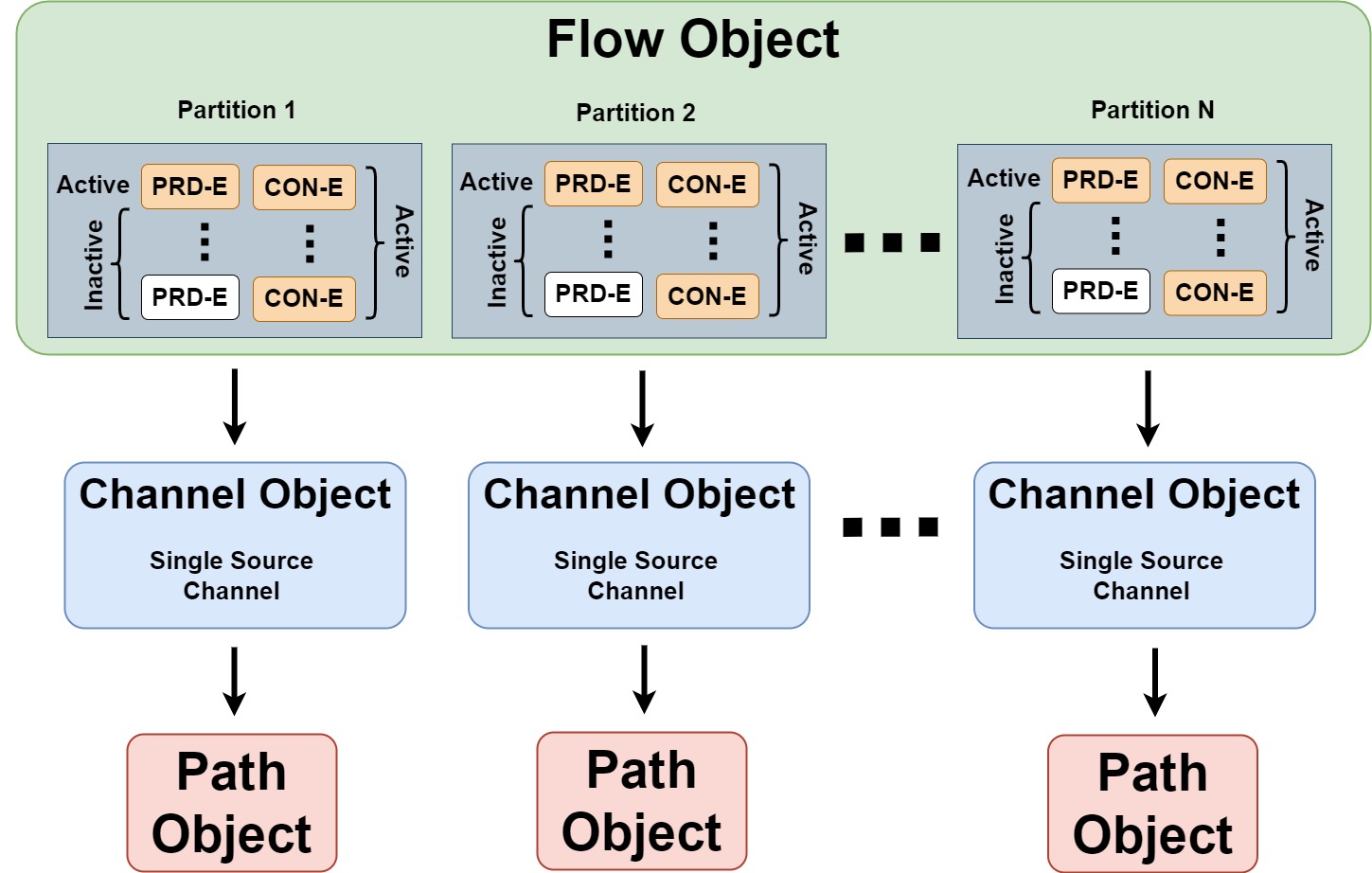}\bf\color{DarkOrchid}
    \centering
    \caption{\color{Black}\small Flow, Channel, and Path objects in Yodel-SLSM/Yodel-SLAC services (PRD-E=Producer Edge and CON-E=Consumer Edge).}
\end{figure}

According to its service model, a Yodel-SLSM/Yodel-SLAC Flow object can contain multiple data producer and consumer Edge Nodes that are organized in one or more partitions that can change dynamically. Within each partition, there is only one active data producer Edge Node, as shown in Figure 21. Separate partitions are also assigned separate single source Channel objects and IDs. Hence, the service Flow object may have one or more Channel objects, depending on the number of the Community's partition. We note that there will be a separate Path object for each partition in the Community, as also shown in Figure 21.

\color{DarkOrchid} \subsubsection{Relevant Use Cases}\color{Black}

The Yodel-SLSM and Yodel-SLAC multicast services are ideal services for enabling load balancing and reducing communication latency in multicast application environments. For instance, in service and micro-service environments multiple service replica/instance may exist where each service replica/instance can process events and transactions in the system independently of others. Using any of the Yodel-SLSM and Yodel-SLAC services, micro-service environments can benefit from load balancing within individual Communities to improve service response time (i.e., lower communication latency), and better utilization of the service processing capabilities. The Yodel-SLSM and Yodel-SLAC services are also capable of providing content delivery networks with native support for regional content distribution while supporting them with network-level multicast capabilities. In this scenario, content delivery networks can use separate partitions to distribute their contents to the content consumers/recipients regionally. Partitions can then be created based on the availability of the content producers in the network. We note that since partitioning can change dynamically, the Yodel-SLSM and Yodel-SLAC services are capable of responding to changes in the network quickly and can adapt to new network topologies rapidly. Such capability can also improve the content delivery networks response time to the changes in content delivery requests.

\color{DarkOrchid} \subsection{The Yodel Multi Source Multicast (Yodel-MSM) Service}\color{Black}
\color{DarkOrchid} \subsubsection{Key Service Specifications}\color{Black}

Unlike the Yodel-SSM, Yodel-AC, and Yodel-SLSM/SLAC services, the Yodel-MSM service is designed to allow all data producers in a Community to simultaneously send their data messages to all data consumers in that Community. The fundamental assumption here is that different data producers can generate different data objects which the Community's data consumers can receive them all regardless of the their similarities and differences. In terms of service specifications, the Yodel-MSM does not impose any limitations in terms of the number of data producer and consumer Edge Nodes in the Community's Flow object. Hence, the service completely follows the steps defined in section 5. As stated in section 4.4.4, the Yodel-MSM service can be also used in its any cast shape, called the Yodel-MSAC, wherein a randomized or dedicated subset of all data consumers in the Community can receive the Community's data messages as explained in section A.2 of this appendix.

\color{DarkOrchid} \subsubsection{Service Flow Management}\color{Black}

According to the service model, the Flow objects of a Community with Yodel-MSM and Yodel-MSAC services can have multiple data producer and consumer Edge Nodes, as shown in Figure 22. The Community's Flow object can also have a single multi-source Channel object and ID that is composed of multiple data producer and consumer Edge Nodes. Since the Channel object is multi-source, the Channel of a Community with Yodel-MSM and Yodel-MSAC services can have multiple Path objects each originating from a distinct producer Edge Node in the Community, as also shown in Figure 22. We note that each Path object is managed independently of other Paths in the Community.

\begin{figure}[h!]
    \includegraphics[height=6cm,width=5.75cm]{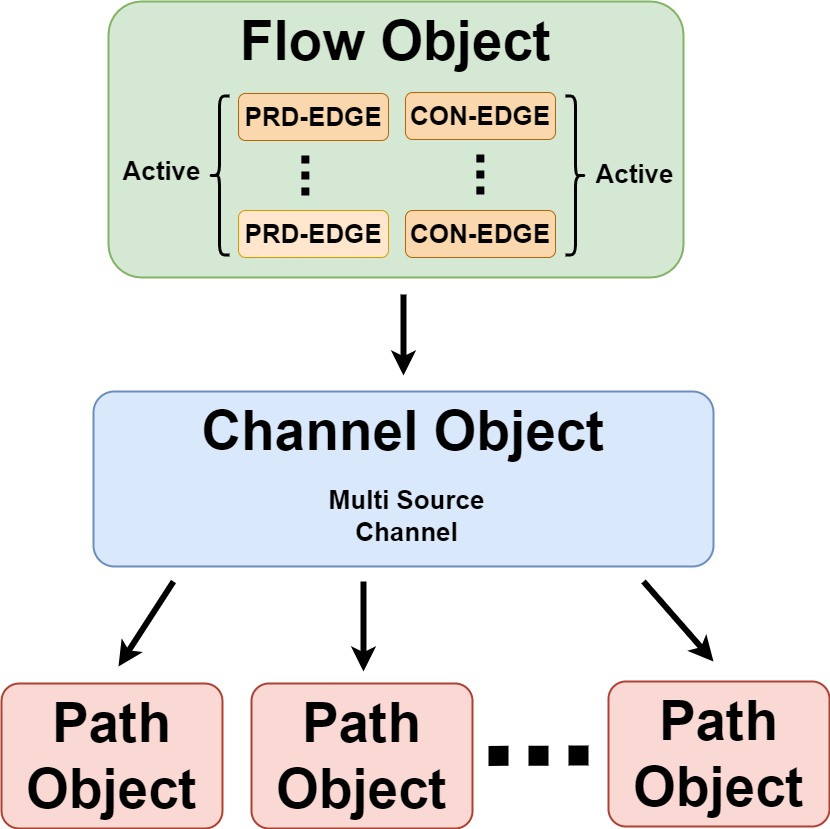}\bf\color{DarkOrchid}
    \centering
    \caption{\color{Black}\small Flow, Channel, and Path objects in Yodel-MSM/Yodel-MSAC services (PRD=Producer and CON=Consumer).}
\end{figure}

\color{DarkOrchid} \subsubsection{Relevant Use Cases}\color{Black}

Given the highly flexible nature of the Yodel-MSM and Yodel-MSAC services, these multicast service models can be used in variety of scenarios that involve the collection and distribution of events and telemetry data. For instance, infrastructure monitoring is a common use case in many Internet of Things applications. Smart grids, smart homes, and smart transportation systems are all examples of such IoT applications where a group of sensors and IoT measurement devices are used to monitor the infrastructure condition. These devices can continuously measure, collect, and send their data to a group of data consumers/recipients that process these data to make predictions and build automation. In these scenarios, the Yodel-MSM/MSAC services can provide a very flexible network-level multicast support that can meet the requirements of these IoT applications at multi-domain scales. The Yodel-MSM/MSAC multi-source services can be also used to meet the multicast requirements of many Web and Mobile applications like online messaging, social media, and push/notification services where users can receive all the messages and communications that are generated for them within the applications from all possible sources. For instance, online messaging and social media applications allow users to create groups that they can privately communicate with each other. Within this groups, all users can simultaneously send their data to the group where all other users (except the sender) can receive them instantly. Although, such communications usually happens by the aid of a messaging broker server, Yodel-MSM/MSAC can provide the applications with the opportunity to instead use the networking multicast services. 

\color{DarkOrchid} \subsection{The Yodel Many-to-Many Multicast (Yodel-MMM) Service}\color{Black}
\color{DarkOrchid} \subsubsection{Key Service Specifications}\color{Black}

The Yodel-MMM service also called the Yodel-3M is the most versatile Yodel multicast service, wherein every Community member/participant can send its data messages to all other members/participants in the Community. In other words, every member receives all other members data messages while sending its own data to those Community members concurrently. In this service, the notion of the \say{Community member/participant} is used to refer to the combined role of data production and consumption in the Community. Therefore, every process in the Community running in a Host is a member/participant and every Yodel Edge Node in the Community is a member/participant Edge Node. Joining a Community with Yodel-MMM service model happens the same way as it is described in section 5.1 with a small difference that once a member/participant join the Community, the Community's Channel ID will be recorded on both the PCT and PPT tables of the corresponding Edge Node to denote both the data producer and consumer roles for the Host where the \say{about-to-join} member/participant is running as well as the PRT and CRT tables on the given Host. The Channel ID will be also shared with the member/participant during the join process which enables the given member/participant to both produce and consume data messages in the Community. The corresponding Edge Node also becomes both a producer and a consumer Edge Node in the Community's Flow object. In all other aspects, the Yodel-MMM service follows the same steps we described in section 5 to enable Community members to interact.

\color{DarkOrchid} \subsubsection{Service Flow Management}\color{Black}

\begin{figure}[h!]
    \includegraphics[height=6cm,width=5.75cm]{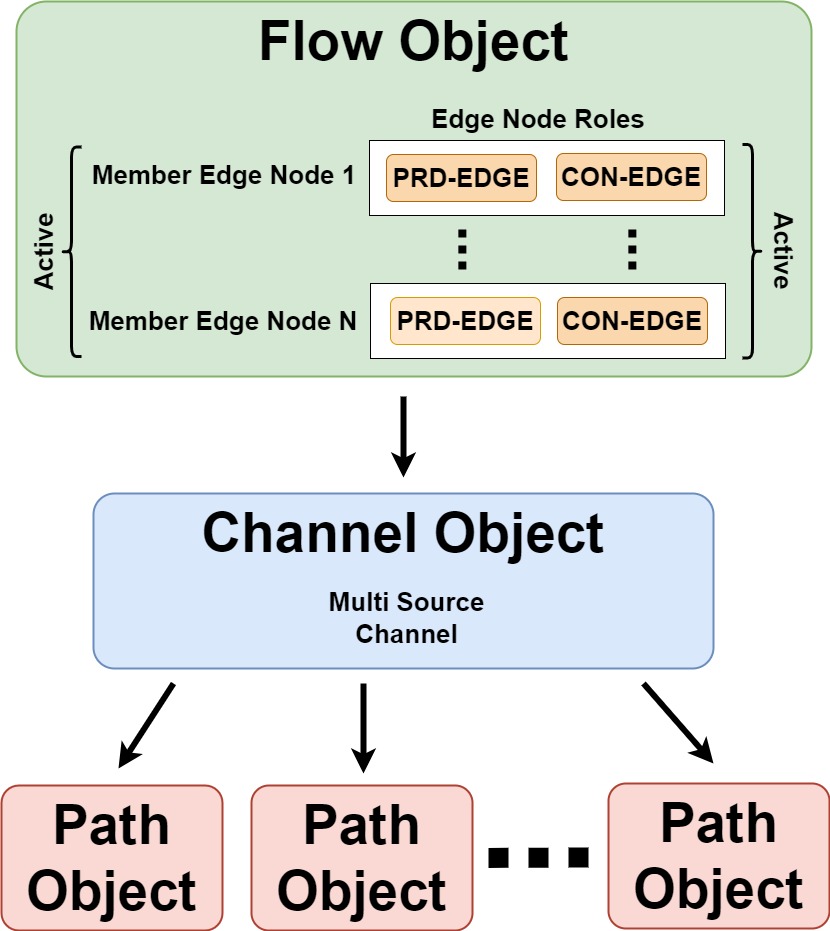}\bf\color{DarkOrchid}
    \centering
    \caption{\color{Black}\small Flow, Channel, and Path objects in Yodel-MMM service (PRD=Producer and CON=Consumer).}
\end{figure}

Similar to the Yodel-MSM/MSAC services, the Yodel-MMM service does not impose any constraints regarding the number of possible Edge Nodes in the Community's Flow object. The only fundamental difference between the Yodel-MMM service and other Yodel multicast services is that every member/participant Edge Node in the Community is considered as both a producer and a consumer Edge Node once it joins the Community, as shown in Figure 23. In other words, a single attempt to join the Community is enough for an Edge Node to obtain both producer and consumer Edge roles whereas in other Yodel multicast services an Edge Node can be a producer and consumer Edge in the Community depending on whether there are data producers and consumers in the Community that are connected to the network through the given Edge Node. Much like the Yodel-MSM/MSAC services, a Community with the Yodel-MMM multicast service model has a single multi-source Channel object which can spawn into one or more Path objects depending on the number of member/participant Edge Nodes in the Community. We note that since each Edge Node in the Community can play both the producer and consumer Edge roles, the number of Path objects in the Community is equivalent to the number of Community's participant Edges.

\color{DarkOrchid} \subsubsection{Relevant Use Cases}\color{Black}

Nowadays, and specially after the COVID-19 outbreak, online communication tools (e.g., online meeting applications) have become very popular for attending meetings remotely, participating in events, offering telemedicine services, or even presenting in a conference held across the globe. Such online remote communication tools typically allow users in an online meeting spanning across the globe to communicate simultaneously. Several key technologies (e.g., multicasting and edge and cloud computing) are commonly involved in designing such systems to ensure high-availability and to provide a reliable user experience. That being said, online communication tools can hugely benefit from network-level multicast support to meet their standards where we believe that the Yodel-MMM service model can provide extensive multicast support for enabling communication in multi-domain settings.